\documentclass[aps,prd,twocolumn,superscriptaddress,nofootinbib, floatfix,10pt]{revtex4-2}
\usepackage[utf8]{inputenc}
\usepackage{amsmath,amssymb,bm,graphicx}
\usepackage{multirow}
\usepackage{xcolor}
\definecolor{lilac}{HTML}{8545f4}
\usepackage[
    colorlinks=true,
    linkcolor=lilac,
    citecolor=lilac,
    urlcolor=lilac
]{hyperref}

\newcommand{\Pt}{\mathcal P_t}
\newcommand{\Ps}{\mathcal P_s}
\newcommand{\F}{\mathcal F}
\newcommand{\Mpc}{\mathrm{Mpc}}
\newcommand{\LCDM}{\Lambda\mathrm{CDM}}
\newcommand{\Sr}{S_r}
\newcommand{\Snt}{S_{n_t}}

\begin{document}

\title{Blue-tilted primordial gravitational waves: model landscape and CMB forecasts}

\author{Francesca \surname{Di Nella}}
\email{francesca.dinella@uniroma1.it}
\affiliation{Dipartimento di Fisica and INFN, Sapienza Universit\`a di Roma, Piazzale Aldo Moro 5, 00185 Rome, Italy}
\affiliation{Dipartimento di Fisica, Universit\`a di Roma Tor Vergata, Via della Ricerca Scientifica 1, 00133 Rome, Italy}

\author{Alessandro Melchiorri}
\affiliation{Dipartimento di Fisica and INFN, Sapienza Universit\`a di Roma, Piazzale Aldo Moro 5, 00185 Rome, Italy}

\date{\today}

\begin{abstract}
A blue primordial tensor spectrum, $n_t>0$, is incompatible with the leading-order consistency relation $n_t=-r/8$ of canonical single-field slow-roll inflation. We review representative mechanisms that can produce blue tensor spectra and investigate their CMB detectability with an idealized \textit{LiteBIRD}+\textit{Planck} configuration using a Fisher forecast. We evaluate the tensor amplitude and tilt on a $22\times22$ grid in $(n_t,r_{0.002})$, initially neglecting foreground residuals and instrumental systematics, and then testing the results against a simple foreground residual model and delensing. Letting $n_t$ be a free parameter significantly degrades the constraint on the tensor normalization at $k_*=0.002\,\Mpc^{-1}$, and most of this degradation comes from the tilt itself rather than from the marginalization over the six $\LCDM$ parameters. $B$-mode polarization provides most of the sensitivity: temperature and $E$-mode spectra alone reach $r/\sigma_r=3-5$ only for extremely blue, large-amplitude spectra. The differences between $TE+BB$ and $EE+BB$ reflect their constraints on the scalar parameters: with \textit{Planck} priors or fixed scalar parameters the two subsets perform equally. A $20\%$ reduction of the lensing power in the $B$-mode spectrum through a phenomenological prescription improves $\sigma_r$ by up to about $17\%$. For our baseline analysis in the full configuration, requiring both $r/\sigma_r>3$ and $n_t/\sigma_{n_t}>2$ selects over one third of the grid, with the smallest tilt satisfying both criteria being $n_t\simeq0.17$ at $r_{0.002}\simeq2.6\times10^{-2}$. Including foreground residuals shrinks this region mostly at $n_t<1.5$, while delensing largely recovers the loss. Measuring the string-gas tilt $n_t\simeq0.034$ would require nearly complete delensing and about five times the polarization sensitivity of \textit{LiteBIRD}. Present bounds, with the power law extrapolated to interferometer frequencies, leave about $4\%$ of the characterization region in the baseline case, and about $1\%$ with a foreground residual $\epsilon=0.01$ and delensing with external tracers. Mapped back onto the model landscape, this singles out mildly blue models with a non-negligible large-scale amplitude as the CMB-relevant targets.
\end{abstract}

\maketitle

\section{Introduction}\label{sec:introduction}

Primordial gravitational waves (GWs) are a generic prediction of inflation and a direct probe of the early Universe \cite{Grishchuk1975,Starobinsky1979,Maggiore2000}. In the standard calculation, vacuum fluctuations of the metric are amplified during inflation with an amplitude set by the expansion rate. For canonical single-field slow-roll inflation with Einstein gravity and a Bunch-Davies initial state, the tensor spectrum is nearly scale invariant and slightly red, and obeys the leading-order consistency relation $n_t=-r/8$. Detecting this background through the $B$-mode polarization of the Cosmic Microwave Background (CMB) would provide a definitive smoking-gun signature of inflation, making it one of the central goals of modern observational cosmology \cite{Planck2018X,BICEPKeck2021}.

A blue spectrum, ($n_t>0$), requires departures from the canonical assumptions. Possible mechanisms include violation of the null energy condition (NEC), modified tensor dynamics, nonstandard initial states, and additional sources of tensor perturbations \cite{Wang2014}. If maintained over a sufficiently broad range of scales, a blue spectrum also enhances the stochastic background in the frequencies probed by pulsar-timing arrays (PTAs) \cite{Foster1990}, LISA \cite{LISA}, the Einstein Telescope \cite{ET}, and Cosmic Explorer \cite{CE}. Recent evidence for a nanohertz background \cite{NANOGrav15} has renewed interest in such mechanisms, although its primordial origin remains uncertain \cite{NANOGravNP2023}.
Joint interpretations of CMB, PTA, and interferometer observations depend on the spectral shape assumed across the full frequency range. In particular, recent analyses find that an unbroken power law does not generally provide a satisfactory description, favoring spectral breaks or non-primordial source populations \cite{Galloni2023,Vagnozzi2023}.

Constraining the tensor tilt is difficult when the amplitude is small. Published CMB and multiband bounds depend on the prior choices and, in multiband analyses, on the extrapolation across many decades in frequency \cite{Meerburg2015,Lasky2016,Cabass2016,Planck2018X, Galloni2023,Galloni2024,Tristram2022,LVK2021}. Previous forecasts have studied the sensitivity to the tensor tilt and its dependence on foregrounds and experimental design \cite{Wu2014,Huang2015,Campeti2021}.

Here we map this sensitivity across a two-dimensional grid of fiducial amplitudes and tilts, compare four combinations of CMB spectra with the same treatment of the cosmological parameters, and relate the resulting trends to representative production mechanisms. We use an idealized \textit{LiteBIRD}+\textit{Planck} configuration to combine sensitive, large-scale polarization measurements with high-resolution observations on smaller scales. Our baseline neglects foreground residuals and instrumental systematics, so it represents statistical sensitivity rather than a full mission forecast. We then test how far that baseline moves when a simple foreground residual is added to the $B$-mode covariance, when the scalar parameters are fixed or constrained by \textit{Planck} priors, and when lensing is reduced with a constant factor or with specific tracers. We also ask what it would take to measure the small tilt predicted by string-gas cosmology, and which part of the plane is still allowed by present data.

Section~\ref{sec:spectrum} defines our spectral conventions and the integrated bound on the GW energy density. Sections~\ref{sec:landscape} and \ref{sec:plane} review the model landscape and the limitations of the $(n_t,r)$ parametrization. Section~\ref{sec:method} describes the forecast method, Sec.~\ref{sec:results} presents and discusses the results, and Sec.~\ref{sec:conclusions} summarizes our conclusions. Appendix~\ref{app:statistics} derives the statistical relations used in the text and describes a likelihood-based test of the Gaussian approximation, Appendix~\ref{app:foregrounds} details the foreground model, and Appendix~\ref{app:delensing} the delensing with specific tracers.

\section{Tensor spectrum and consistency relation}\label{sec:spectrum}

We define the dimensionless tensor power spectrum as
\begin{equation}
\Pt(k)=A_t\exp\!\left[n_t x+\tfrac12\alpha_t x^2+\cdots\right],\qquad x\equiv\ln(k/k_*),
\label{eq:spectrum}
\end{equation}
where $n_t=d\ln\Pt/d\ln k|_{k_*}$ and $\alpha_t=dn_t/d\ln k|_{k_*}$ are the tensor tilt and its running at the pivot scale $k_*$. The tensor-to-scalar ratio is $r(k_*)=\Pt(k_*)/\Ps(k_*)$, where $\Ps$ is the dimensionless power spectrum of the comoving curvature perturbation, with scalar tilt $n_s$. When both runnings vanish, we can relate two tensor-to-scalar ratios computed at different comoving wavenumbers as
\begin{equation}
r(k_2)=r(k_1)\left(\frac{k_2}{k_1}\right)^{n_t-(n_s-1)}.\label{eq:pivot}
\end{equation}
Values of $r$ quoted at different pivots therefore cannot be compared without specifying both tilts. For the two pivots used in this work, $k_*=0.002\,\mathrm{Mpc}^{-1}$ and $k_s=0.05\,\mathrm{Mpc}^{-1}$, the conversion factor is $25^{\,n_t-(n_s-1)}$, which exceeds $10^{4}$ for $n_t=3$.

In canonical slow-roll inflation, $n_t=-2\epsilon_1$ and $r=16\epsilon_1$ at leading order, with $\epsilon_1=-\dot H/H^2$, which gives the consistency relation \cite{Baumann2009}
\begin{equation}
n_t=-\frac{r}{8}.\label{eq:consistency}
\end{equation}
In a spatially flat, expanding universe governed by Einstein gravity, the NEC implies $\epsilon_1\geq0$. Within the standard leading-order calculation, a blue tilt requires $\dot H>0$, or a modification of the tensor dynamics, of the initial state, or of the sources. Modifying the scalar kinetic term alone does not change the sign of the tensor tilt as long as the NEC holds.

A positive tilt has a direct impact on the present-day energy density of the primordial gravitational-wave background. For modes well inside the horizon today, the GW energy density per logarithmic wavenumber is \cite{Caprini2018,WatanabeKomatsu2006}
\begin{equation}
\Omega_{\rm GW,0}(k)=\frac{1}{12}\left(\frac{ck}{a_0H_0}\right)^2
\Pt(k)\,\overline{T_h^2(k,\eta_0)},\label{eq:omega}
\end{equation}
where $T_h$ is the transfer function of the tensor amplitude, $\eta_0$ is the present conformal time, and the overline denotes an average over oscillations. Defining the observed frequency as $f=ck/(2\pi a_0)$, modes entering the horizon during radiation domination inherit a primordial power-law spectrum, yielding approximately $\Omega_{\rm GW,0}\propto f^{n_t}$, up to corrections from the evolution of the number of relativistic degrees of freedom and from damping effects \cite{Caprini2018,WatanabeKomatsu2006}.

The growth of the GW energy density with frequency implied by a blue tilt cannot continue indefinitely: Big-bang nucleosynthesis (BBN) and CMB measurements constrain the total energy density in primordial gravitational waves through the effective number of relativistic species \cite{Maggiore2000,Caprini2018}
\begin{equation}
\int_{f_{\rm low}}^{f_{\rm UV}}d\ln f\,
\Omega_{\rm GW,0}(f)h^2
\lesssim5.6\times10^{-6}\,\Delta N_{\rm eff}.\label{eq:bbn}
\end{equation}
The lower limit corresponds to modes inside the horizon at BBN, while the ultraviolet cutoff $f_{\rm UV}$ depends on the production mechanism and on the subsequent evolution. A quantitative exclusion requires specifying the adopted limit on $\Delta N_{\rm eff}$. For sufficiently blue spectra, the high-frequency growth must eventually be interrupted by a turnover or cutoff. The resulting constraint on $(n_t,r)$ depends on the tensor amplitude, frequency range, and thermal history \cite{Kuroyanagi2015,Giare2023}.

\section{Mechanisms for blue tensor spectra}\label{sec:landscape}

\begin{figure*}[t]
\centering
\includegraphics[width=\textwidth]{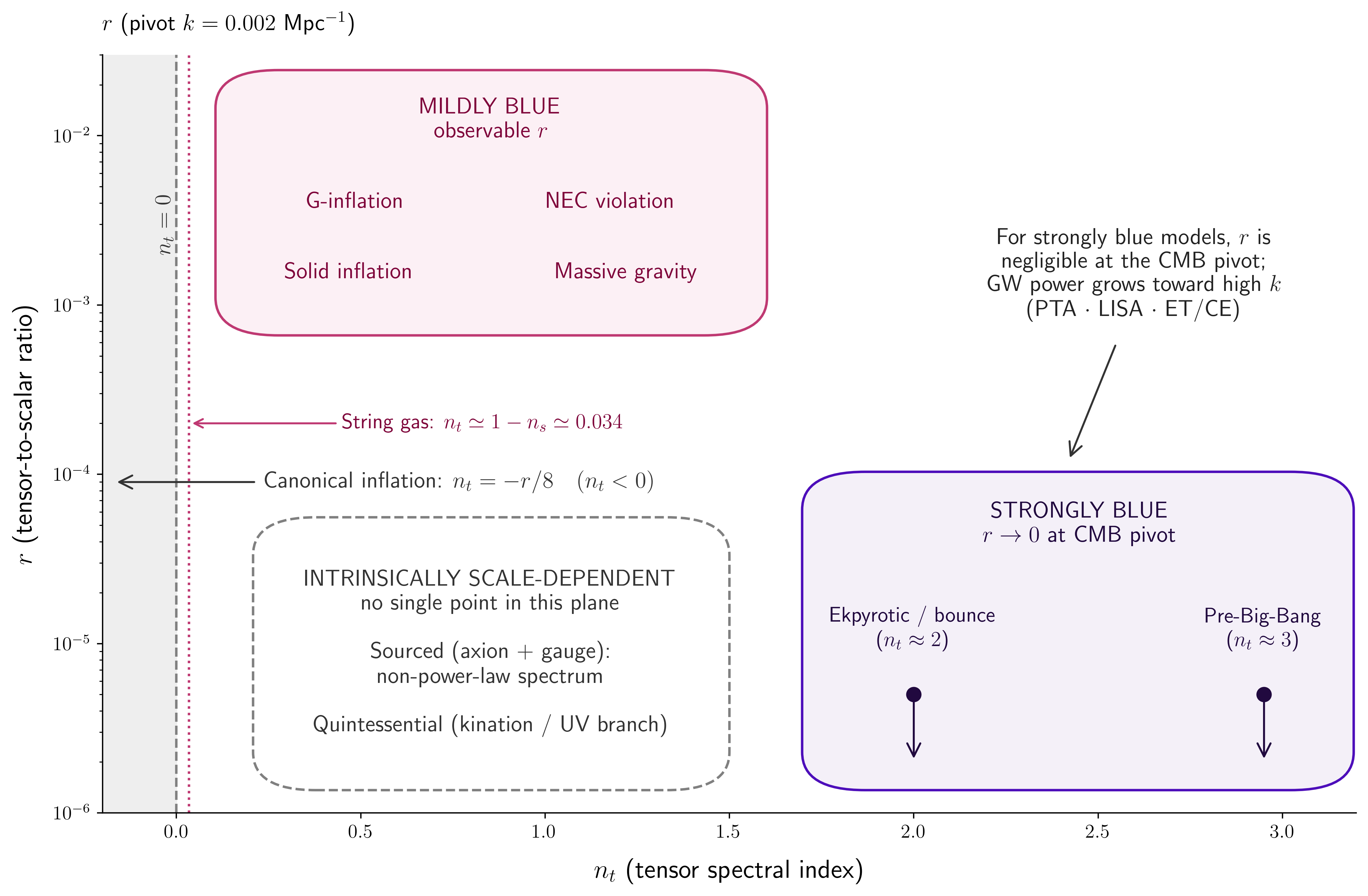}
\caption{Schematic and qualitative classification of blue-tilted models, grouped by the interpretation of their spectra. The dashed vertical line at $n_t=0$ separates canonical red-tilted inflation ($n_t=-r/8$, arrow) from the blue region. The dotted line, at $n_t\simeq0.034$, marks the near-Hagedorn string-gas estimate for our fiducial $n_s$. Mildly blue models exhibit potentially observable CMB amplitudes, whereas strongly blue models feature $r\to0$ at the CMB pivot, with downward arrows indicating that their amplitudes can extend far below our grid. Intrinsically scale-dependent scenarios do not map onto a single point and are grouped in a dashed box. The locations are illustrative and indicate only representative orders of magnitude.}
\label{fig:landscape}
\end{figure*}

We group scenarios with $n_t>0$ into three broad families according to their location in the $(n_t,r)$ plane. The classification is summarized in Fig.~\ref{fig:landscape}. Individual model realizations do not populate arbitrary pairs of $(n_t,r)$ because their physical mechanisms tie together their amplitude and slope, making them testable via existing CMB and multi-band constraints \cite{Benetti2021,Vagnozzi2023,NANOGravNP2023}. In our forecast, the CMB-scale normalization and tilt are treated as phenomenological parameters.

\subsection{Models with a potentially observable CMB amplitude}
\textit{NEC violation and generalized inflation.} G-inflation and its generalizations \cite{Kobayashi2010,Kobayashi2011}, motivated by Galileon interactions \cite{Nicolis2009}, admit stable NEC-violating backgrounds in suitable regions of parameter space. If the tensor normalization and propagation remain approximately standard, $\epsilon_1<0$ yields a blue tensor tilt. More generally, the tensor spectrum can be modified by NEC violation \cite{Baldi2005,Creminelli2010}, by a time-dependent tensor sound speed \cite{Cai2016}, or by higher-curvature interactions \cite{Giare2021}. Amplitude and tilt depend on the specific realization and can fall within the range accessible to CMB experiments.

\textit{Solid inflation.} In solid inflation, the symmetry-breaking pattern of the inflationary medium modifies the tensor evolution and allows a blue spectrum \cite{Endlich2013}. The tensor-to-scalar ratio can be suppressed relative to standard inflation while remaining potentially observable.

\textit{Massive tensors and broken spatial symmetries.} An effective graviton mass during inflation \cite{Fujita2019} or broken spatial diffeomorphisms \cite{Cannone2015} can generate a blue tilt, with amplitude and slope determined by the additional mass scales and by the background evolution.

\textit{String-gas cosmology.} Thermal fluctuations of a string gas in a quasistatic Hagedorn phase \cite{BrandenbergerVafa1989} generate nearly scale-invariant scalar perturbations with a small red tilt \cite{Nayeri2006}, while anisotropic-pressure fluctuations source a slightly blue tensor spectrum \cite{Brandenberger2007}. In the usual near-Hagedorn approximation, and neglecting subleading logarithmic corrections, the two tilts are related by \cite{Brandenberger2015}
\begin{equation}
n_t\simeq1-n_s\simeq0.034,\label{eq:stringgas}
\end{equation}
where the numerical value corresponds to our fiducial $n_s$. The tensor amplitude depends on the dynamics of the exit from the Hagedorn phase and is not a model-independent prediction, but it is expected to be larger than in simple slow-roll scenarios \cite{Brandenberger2007}. Observational constraints on this scenario were discussed in Ref.~\cite{StewartBrandenberger2008}.

\subsection{Steep spectra with suppressed large-scale tensors}
\textit{Pre-big-bang cosmology.} The dilaton-driven phase produces a steep vacuum tensor spectrum, with $n_t\simeq3$ up to logarithmic corrections \cite{GasperiniVeneziano1993,Gasperini2003}. Its normalization and high-frequency branches depend on the subsequent evolution, while its amplitude on CMB scales is generally extremely small.

\textit{Ekpyrotic contraction.} The vacuum tensor spectrum generated during ekpyrotic contraction is strongly blue, $n_t\simeq2$, with a very small amplitude on CMB scales \cite{Khoury2001,Boyle2004}.

Because typical realizations of these scenarios usually predict CMB amplitudes far below our grid, they are not the direct targets of this forecast, and evaluating their detectability at higher frequencies requires specifying their high-energy completion, spectral breaks, and transfer function.

\subsection{Sourced and transfer-induced blue branches}
Gauge fields coupled to the inflaton or to a spectator axion can source tensor modes with strong scale dependence, non-Gaussianity, and chirality \cite{Sorbo2011,BarnabyPeloso2011,Cook2012,Barnaby2012,Mukohyama2014,Namba2016}. A local power law can describe a smooth portion of such spectra, but sharp features cannot be captured by the simple power-law parametrization used here. Scalar-induced GWs \cite{Domenech2021} and GWs from preheating \cite{KhlebnikovTkachev1997,EastherLim2006} likewise rise around a characteristic production scale instead of following a single universal power law.

Quintessential inflation \cite{PeeblesVilenkin1999} illustrates a different mechanism, where a post-inflationary kination phase ($w \simeq 1$) boosts high-frequency gravitational wave energy density via the transfer function while leaving the CMB-scale primordial spectrum approximately as in standard inflation. Because these complex spectra do not map cleanly onto the $(n_t,r)$ plane as power-law realizations, they lie outside our forecast scope. Furthermore, our Fisher analysis assumes a Gaussian, parity-even covariance and thus neglects the higher-order and parity-odd information carried by sourced, chiral signals.

\section{The \texorpdfstring{$(n_t,r)$}{(nt,r)} plane and its limitations}\label{sec:plane}

There are three main limitations to representing these models in the $(n_t,r)$ plane.

\textit{Pivot dependence.} For a pure power law, $n_t$ is independent of the pivot, whereas $r$ changes according to Eq.~\eqref{eq:pivot}. A quoted value of $r$ therefore requires the pivot and the scalar convention to be specified. Once a non-zero running $\alpha_t$ is allowed, $n_t$ also becomes pivot dependent.

\textit{Range of scales.} The CMB probes a limited range of comoving wavenumbers, and a power law valid at CMB scales is not guaranteed to hold in the PTA or interferometer bands: peaked sources, spectral breaks, and high-frequency cutoffs must be treated explicitly when extending forecasts to multi-band experiments. Relating the present-day GW energy-density slope to the primordial tensor tilt also requires the appropriate transfer function.

\textit{Propagation and high-frequency extrapolation.} Our CMB transfer functions assume standard tensor propagation in general relativity and the scalar background of Sec.~\ref{sec:method}. Models with a modified graviton mass or propagation speed can be represented in the $(n_t,r)$ plane only if these effects are negligible by the time the relevant tensor modes re-enter the horizon. Persistent propagation effects, scalar-tensor correlations, and parity-odd or non-Gaussian signals are not included in the forecast.

The effect of extrapolating across the $\sim 20$ orders of magnitude separating the CMB ($f_* \sim 10^{-18}\,\mathrm{Hz}$) from ground-based interferometers ($f \sim 10^2\,\mathrm{Hz}$) can be shown by including a constant running,
\begin{equation}
\ln\left[\frac{\Pt(k)}{A_t}\right]=n_tx+\tfrac12\alpha_t x^2.\label{eq:running}
\end{equation}
While the CMB window extends at most to $k=0.1\,\Mpc^{-1}$ ($x\simeq3.91$), ground-based interferometer bands correspond to $x\simeq44$. A modest running of $\alpha_t=-0.01$ produces a percent-level suppression at the smallest CMB scales, but suppresses the constant-tilt extrapolation by a factor $\simeq10^{-4}$ in interferometer bands.

\begin{figure}[t]
\centering
\includegraphics[width=\columnwidth]{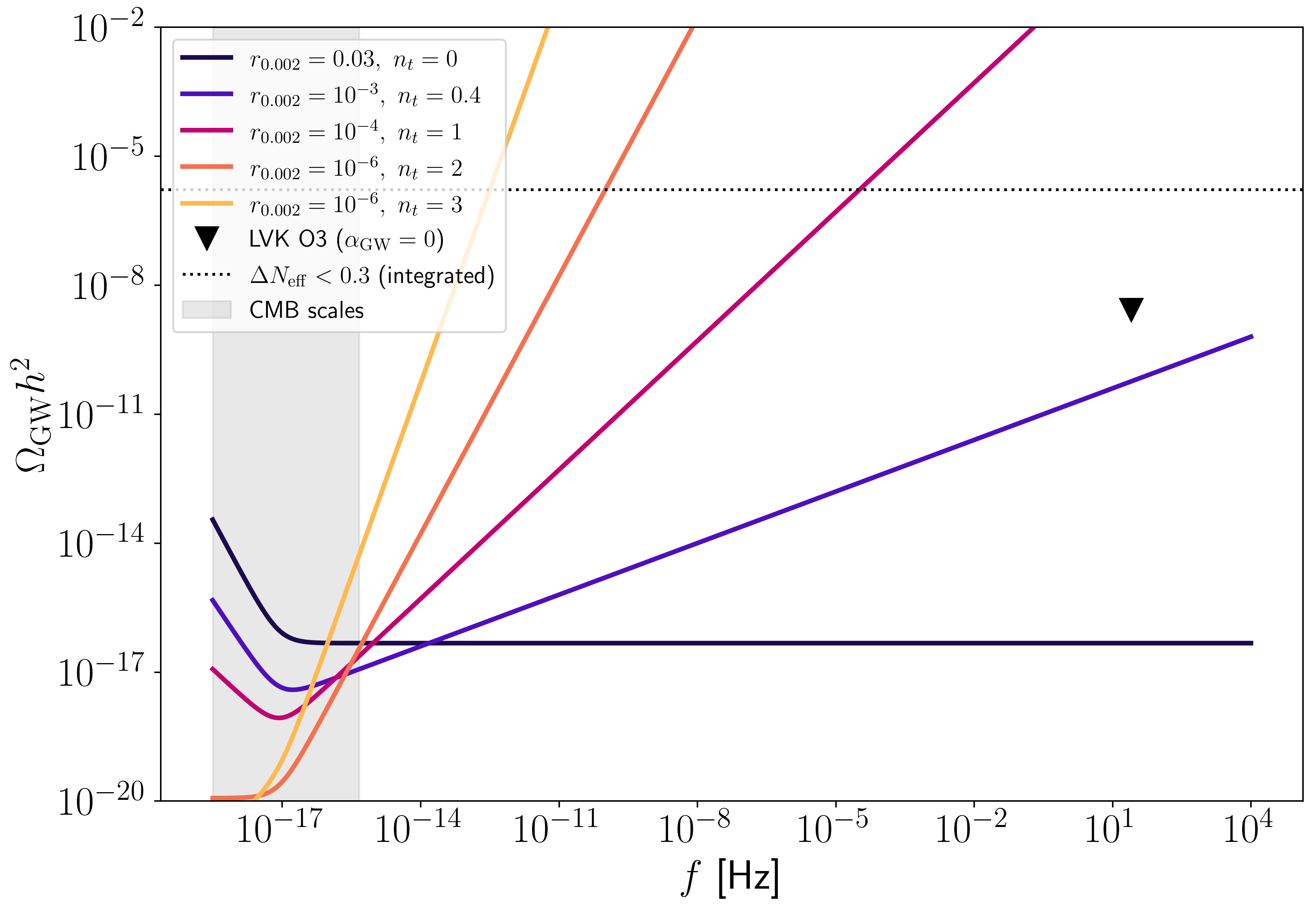}
\caption{Present-day $\Omega_{\rm GW}h^2$ for power-law extrapolations of representative grid points, starting at the lowest wavenumber probed by the CMB (grey band). The triangle is the LVK O3 upper limit at $25\,$Hz for a flat spectrum and the dotted line the integrated nucleosynthesis bound with $\Delta N_{\rm eff}<0.3$ (Sec.~\ref{sec:constraints}).}\label{fig:omega}
\end{figure}

Equation~\eqref{eq:running} describes the primordial spectrum while Equation~\eqref{eq:omega} turns it into the present-day energy density through the transfer function. Each tensor mode follows $h''+2(a'/a)h'+k^2h=0$ in conformal time and is frozen outside the horizon, so that $T_h\to1$ for $k\eta\ll1$ \cite{WatanabeKomatsu2006}. For a mode that enters during radiation domination, we find $T_h=j_0(k\eta)=\sin(k\eta)/(k\eta)$: its envelope decays as $a_k/a$, where $a_k$ is the scale factor at horizon entry, $k=a_kH(a_k)$, and the oscillation average is $\overline{T_h^2}=\tfrac12(a_k/a_0)^2$. Modes that enter during matter domination have $T_h=3j_1(k\eta)/(k\eta)$ and add a term proportional to $(k_{\rm eq}/k)^2$. Joining the two limits and including the change in the number of relativistic species gives \cite{Lasky2016,WatanabeKomatsu2006}
\begin{equation}
\begin{split}
\Omega_{\rm GW,0}h^2&=\mathcal W(x)\,\Pt(k),\\
\mathcal W(x)&=\frac{\Omega_rh^2}{24}\left[1+\frac{9}{32}\left(\frac{k_{\rm eq}}{k}\right)^2\right]
\frac{g_*}{g_{*0}}\left(\frac{g_{*s0}}{g_{*s}}\right)^{4/3}.
\end{split}\label{eq:weight}
\end{equation}
The present-day slope equals the primordial tilt, $d\ln\Omega_{\rm GW,0}/d\ln f=n_t$, as long as the Universe is radiation dominated when the relevant modes enter and the number of relativistic species does not change in between. We neglect neutrino free streaming, which lowers $\Omega_{\rm GW,0}$ by $35.5\%$ for $10^{-16}<f<2\times10^{-10}\,$Hz \cite{WatanabeKomatsu2006} and leaves interferometer frequencies untouched. Figure~\ref{fig:omega} shows the effect of the extrapolation on today's GW energy density at selected points of our grid, along with current BBN and interferometer bounds.

Introducing negative running to force a spectral turnover does not automatically satisfy the BBN bound. While $\alpha_t<0$ produces a primordial peak at $x_{\rm pk}=n_t/|\alpha_t|$,
the BBN bound constrains the total integrated energy density
\begin{equation}
I_{\rm GW}=A_t\int_{x_{\rm low}}^{x_{\rm UV}}dx\,
\mathcal W(x)\exp\!\left(n_tx+\tfrac12\alpha_t x^2\right),\label{eq:integral}
\end{equation}
where $\mathcal W(x)$ includes the transfer function and the conversion to $\Omega_{\rm GW,0}h^2$ of Eq.~\eqref{eq:weight}. Because the contribution beyond the peak must be included, simply bending the spectrum does not guarantee the integrated power remains within observational limits.

Consequently, there is no universal, amplitude-independent ``critical running'' that guarantees consistency with the BBN bound. The result depends on $A_t$, the ultraviolet cutoff, and the transfer function entering $\mathcal W(x)$.
Nonstandard reheating, entropy production, and broken spectra can also suppress the high-frequency signal \cite{Kuroyanagi2015,Gasperini2003}; on the other hand, kination can enhance it. We consequently interpret the grid only as a map of CMB-scale sensitivity.

\section{Forecasting method}\label{sec:method}

\subsection{Fiducial cosmology}\label{sec:fiducial}
We adopt the flat $\LCDM$ scalar cosmology of Table~\ref{tab:fiducial}, consistent with \textit{Planck} 2018 results \cite{Planck2020_I}. We compute a CMB Fisher matrix at each point of a $22\times22$ grid in the tensor parameters, logarithmically spaced in $r_{0.002}\in[10^{-6},3\times10^{-2}]$ and linearly spaced in $n_t\in[0,3]$: the tilt is specified independently and not set by the inflationary consistency relation. We choose the upper edge of the amplitude range to reach the current direct CMB bound (Sec.~\ref{sec:constraints}). The scalar and tensor runnings are set to zero.
The scalar amplitude is defined at $k_s=0.05\,\Mpc^{-1}$, whereas $r$ and the tensor normalization refer to $k_*=0.002\,\Mpc^{-1}$, so that the consistent normalization is
\begin{equation}
\Ps(k_*)=A_s(k_s)\left(\frac{k_*}{k_s}\right)^{n_s-1},\quad
A_t=r_{0.002}\,\Ps(k_*),
\label{eq:normalization}
\end{equation}
which gives $\Ps(k_*)\simeq2.344\times10^{-9}$ for our fiducial parameters.
\begin{table}[t]
\caption{Fiducial scalar parameters and tensor grid; $A_s$ is defined at $k_s=0.05\,\Mpc^{-1}$ and $r$ at $k_*=0.002\,\Mpc^{-1}$.}\label{tab:fiducial}
\begin{tabular}{lc}
\hline\hline
Parameter & Fiducial value or range\\
\hline
$\Omega_bh^2$ & 0.022377\\
$\Omega_ch^2$ & 0.12010\\
$100\theta_*$ & 1.04110\\
$\tau$ & 0.0543\\
$n_s$ & 0.96589\\
$\ln[10^{10}A_s(k_s)]$ & 3.0447\\
$r_{0.002}$ & $[10^{-6},3\times10^{-2}]$, 22 values\\
$n_t$ & $[0,3]$, 22 values\\
\hline\hline
\end{tabular}
\end{table}
Theoretical CMB spectra are computed with the Boltzmann solver \texttt{CAMB} \cite{CAMB}.

We consider two parameter sets. In the \emph{fixed-$n_t$} case, the six scalar parameters and $r$ are varied while $n_t$ is held at its fiducial value; in the \emph{free-$n_t$} case, all eight parameters are varied. Unless stated otherwise, we quote free-$n_t$ uncertainties, marginalized over all other parameters, without external priors. Comparing the two cases isolates the cost of leaving the tilt free. The scalar sector will be well constrained through temperature and $E$ modes, so we also repeat the analysis with two other treatments of the six $\LCDM$ parameters: Gaussian priors from the \textit{Planck} 2018 $TT,TE,EE+$low$T+$low$E$+lensing constraints \cite{Planck2018VI}, and the parameters fixed at their fiducial values. In both cases $r$ and $n_t$ remain free.

We define the marginal signal-to-noise ratios (SNRs) for the tensor amplitude and tilt as
\begin{equation}
\Sr=\frac{r_{\rm fid}}{\sigma_r},\qquad
\Snt=\frac{n_{t,\rm fid}}{\sigma_{n_t}}.\label{eq:snr}
\end{equation}
Unless stated otherwise, all improvements and degradations quoted below refer to $\sigma_r$ or $\sigma_{n_t}$. As a first step, we do not mask regions already disfavored by current data, so our results are not to be read as a fraction of viable models; Sec.~\ref{sec:constraints} shows which part of the grid present bounds still allow.

\subsection{Fisher matrix and instrumental noise}
For statistically isotropic, Gaussian, parity-even temperature and polarization fields with uncorrelated temperature and polarization noise, we define the CMB covariance as
\begin{equation}
\widetilde{\bm C}_\ell=
\begin{pmatrix}
C_\ell^{TT}+N_\ell^T & C_\ell^{TE} & 0\\
C_\ell^{TE} & C_\ell^{EE}+N_\ell^P & 0\\
0 & 0 & C_\ell^{BB}+N_\ell^P
\end{pmatrix},\label{eq:covariance}
\end{equation}
where $C_\ell^{XY}$ is the sky signal and $N_\ell^T$ and $N_\ell^P$ are the temperature and polarization noise spectra. The Fisher matrix is \cite{Tegmark1997,Wu2014}
\begin{equation}
\mathcal F_{ij}=\sum_{\ell=2}^{2500}\frac{(2\ell+1)f_{\rm sky}(\ell)}2
\operatorname{Tr}\!\left[
\widetilde{\bm C}_\ell^{-1}\bm C_{\ell,i}
\widetilde{\bm C}_\ell^{-1}\bm C_{\ell,j}\right],\label{eq:fisher}
\end{equation}
where $\bm C_{\ell,i}=\partial\widetilde{\bm C}_\ell/\partial\theta_i$, and the factor $f_{\rm sky}(\ell)$ approximates the reduction in the number of available modes. We take the sky fraction to be $0.7$ up to $\ell=600$, where we use \textit{LiteBIRD}, and $0.65$ above, where we use \textit{Planck} (Table~\ref{tab:noise}). The marginalized uncertainty on parameter $\theta_i$ is $\sigma_i=\sqrt{(\mathcal F^{-1})_{ii}}$.

Across our dense grid, the Fisher matrix provides a computationally efficient and sufficiently accurate tool to map relative sensitivity trends across the $(n_t,r)$ plane. This approach accurately mirrors full Markov Chain Monte Carlo posteriors when parameters are tightly constrained and the fiducial model is sufficiently far from boundaries \cite{Wolz2012,Sellentin2014}. This approximation breaks down near $r=0$, where the posterior is truncated by $r \geq 0$, the tilt becomes unidentifiable, and constraints on an unmeasured tilt depend strongly on the statistical framework used \cite{Galloni2024}; results in this region should therefore be interpreted with caution.

Because primordial tensors are typically red-tilted, default Boltzmann solver settings cut off tensor spectra around $\ell \approx 1500$ by default. To maintain numerical accuracy for steep blue tilts up to small scales, we set \texttt{max\_l\_tensor = 2500} and \texttt{max\_eta\_k\_tensor = 5000} in \texttt{CAMB}.

For independent frequency channels, inverse-variance weighting gives
\begin{equation}
\begin{split}
N_\ell^X&=\left[\sum_\nu(N_{\ell,\nu}^X)^{-1}\right]^{-1},\\
N_{\ell,\nu}^X&=\Delta_{X,\nu}^2
\exp\!\left[\frac{\ell(\ell+1)\theta_{b,\nu}^2}{8\ln2}\right],
\end{split}\label{eq:noise}
\end{equation}
where $X=T,P$, $\Delta_{X,\nu}$ is the white-noise level in $\mu\mathrm K$\,rad, and $\theta_{b,\nu}$ is the full width at half maximum of the Gaussian beam in radians. Table~\ref{tab:noise} lists the instrumental specifications for each frequency channel, together with the multipole range assigned to each experiment. For \textit{LiteBIRD}, we restrict the cosmological analysis to the channels between $100$ and $195\,\mathrm{GHz}$, where CMB emission dominates, reserving the remaining low- and high-frequency channels for foreground cleaning. In fact, a parametric separation of CMB, dust and synchrotron with all fifteen \textit{LiteBIRD} bands \cite{Stompor2009} gives the 40 to 89\,GHz bands a total CMB weight of $-0.05$ at $\ell=80$, and the 68 and 78\,GHz bands individually receive $-0.012$ and $0.001$. To a good approximation, we then assume that they are used to measure synchrotron and dust, while the five bands adopted here carry most of the CMB weight.

Since published noise tables are not yet available for the reformed \textit{LiteBIRD} design \cite{LiteBIRD2026}, we read representative polarization depths for its five bands between 94 and 217\,GHz from Fig. 2 of Ref.~\cite{LiteBIRD2026}, converted them to CMB units and kept the beams of Table~\ref{tab:noise}. The $BB$ noise grows by $16\%$ at $\ell=80$ with respect to the 2023 configuration, and our results shift at the sub-percent level, so the conclusions below do not depend on the design update.

\begin{table*}[t]
\caption{Instrumental specifications adopted in the forecast, based on \textit{Planck} \cite{Planck2020_I} and \textit{LiteBIRD} \cite{LiteBIRD2023}.}\label{tab:noise}
\begin{tabular}{cccccc}
\hline\hline
Experiment & $\ell$ range & $\nu$ [GHz] & $\theta_b$ [arcmin] & $\Delta_T$ [$\mu$K-arcmin] & $\Delta_P$ [$\mu$K-arcmin]\\
\hline
\multirow{3}{*}{\textit{Planck}, $f_{\rm sky}=0.65$} & \multirow{3}{*}{$601\leq\ell\leq2500$}
&100&9.7&77.4&117.6\\
&&143&7.2&33.0&70.2\\
&&217&4.9&46.8&105.0\\
\hline
\multirow{5}{*}{\textit{LiteBIRD}, $f_{\rm sky}=0.7$} & \multirow{5}{*}{$2\leq\ell\leq600$}
&100&37.8&4.64&6.56\\
&&119&33.6&3.24&4.58\\
&&140&30.8&3.39&4.79\\
&&166&28.9&3.94&5.57\\
&&195&28.6&4.14&5.85\\
\hline\hline
\end{tabular}
\end{table*}

\begin{figure}[t]
\centering
\includegraphics[width=\columnwidth]{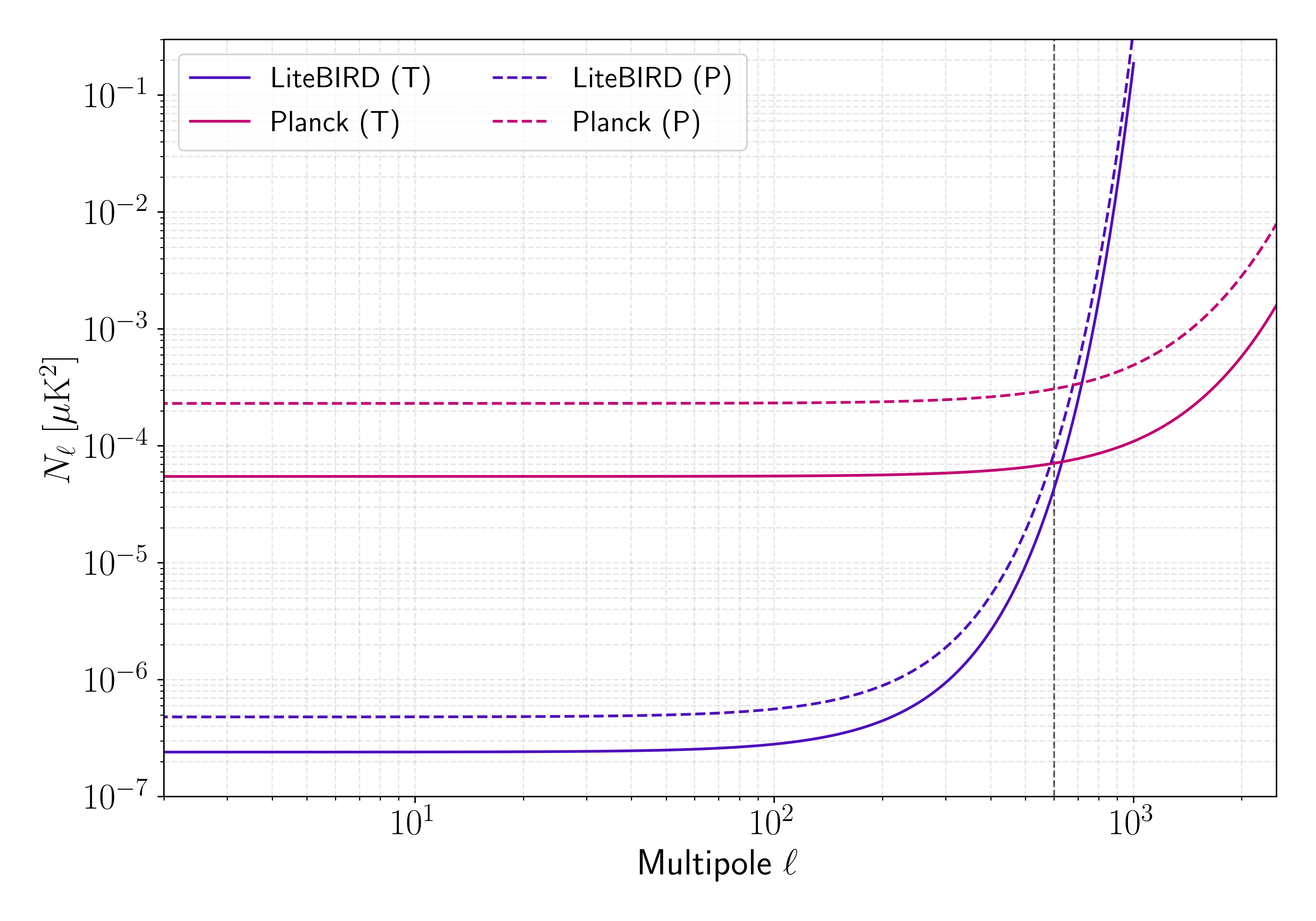}
\caption{Effective noise spectra from Eq.~\eqref{eq:noise} and Table~\ref{tab:noise}. Solid and dashed curves refer to temperature and polarization, respectively. The dotted vertical line marks the switch between \textit{LiteBIRD} and \textit{Planck} at $\ell=600$.}\label{fig:noise}
\end{figure}

The noise curves of the two instruments cross at $\ell\simeq632$ in temperature and $\ell\simeq679$ in polarization (Fig.~\ref{fig:noise}), but we have verified on a coarser grid that moving the transition to either crossing changes the forecast uncertainties at the sub-percent level. We therefore use $\ell=600$ as the boundary between the \textit{LiteBIRD} and \textit{Planck} specifications. Using \textit{Planck} noise at high multipoles provides a conservative baseline that leaves room for further gains from deeper small-scale measurements targeted by next-generation ground-based surveys such as the Simons Observatory \cite{SimonsObs2019}.

Our baseline forecast relies on the optimistic assumption that component separation is efficient enough to reduce foreground residuals below the instrumental noise floor. In a realistic \textit{LiteBIRD} pipeline, residual foreground contamination remains non-negligible and is expected to dominate the uncertainty on $r$ at low amplitudes \cite{LiteBIRD2023}. Galactic dust and synchrotron exhibit steep angular power spectra, with respective $BB$ slopes $C_\ell\propto\ell^{-2.54}$ \cite{Planck2018XI} and $\ell^{-2.22}$ \cite{Martire2022}, indicating that their contribution is strongest at the largest angular scales. Because a blue primordial signal shifts the bulk of the $BB$ signal away from the reionization bump toward the recombination peak and intermediate multipoles, the relative impact of foreground contamination on the overall signal could be less detrimental than in standard red-tilted models. This effect is quantified in Section~\ref{sec:fg} with a simple residual model.

\subsection{Foreground residuals}\label{sec:fg}
To make our forecasts more realistic while avoiding a full component separation, which lies beyond the scope of our work, we model foreground residuals directly in the $BB$ covariance. Co-adding the five \textit{LiteBIRD} channels using the inverse-noise weights built from the channel noise of Eq.~\eqref{eq:noise}, $w_\nu(\ell)=N_{\ell,\nu}^{-1}/\sum_{\nu'}N_{\ell,\nu'}^{-1}$, allows us to evaluate the impact of dust and synchrotron under the conservative assumption that the foregrounds are perfectly coherent across frequencies \cite{TegmarkEisenstein2000}. Given their spectral energy distributions $a_c(\nu)$ in CMB units and angular spectra $C_\ell^c$, the foreground power left in the co-added map is 
\begin{equation}
C_\ell^{\rm FG}=\sum_{c={\rm d,s}}\Big[\sum_\nu w_\nu(\ell)\,a_c(\nu)\Big]^2C_\ell^c .
\end{equation}
A cleaning that leaves a fraction $\epsilon$ of the foreground amplitude leaves $\epsilon^2$ of its power. Earlier forecasts write the residual directly as a fraction of the foreground power in each channel \cite{Verde2006,Creminelli2015}, which corresponds to $\epsilon^2$ here (Appendix~\ref{app:foregrounds} compares the two approaches), so
\begin{equation}
N_\ell^{BB}\ \to\ N_\ell^{BB}+\epsilon^2\,C_\ell^{\rm FG},\qquad 2\leq\ell\leq600 .\label{eq:fgres}
\end{equation}
Dust is a modified black body with the \textit{Planck} 2018 amplitude, slope and spectral parameters for the region LR71 \cite{Planck2018XI}, while synchrotron is a power law with the \textit{Planck} 30\,GHz amplitude and slope of Ref.~\cite{Martire2022} and $\beta_{\rm s}=-3.05$ (full values are listed in Appendix~\ref{app:foregrounds}).  
The residual biases the estimate of $r$ as well as inflating its error but, for simplicity, we assume that the bias is removed and keep only the extra variance.

For our residual toy model, we select benchmark values of $\epsilon = 0.01$ and $0.03$, motivated by \textit{LiteBIRD} sensitivity goals. Specifically, the mission targets a statistical error of $\sigma(r=0)<6\times10^{-4}$ \cite{LiteBIRD2023} and, after the reformation, a total $\sigma_r<10^{-3}$ \cite{LiteBIRD2026}. With our noise and $f_{\rm sky}=0.7$, the first requirement holds for $\epsilon\leq0.029$ and the second for $\epsilon\leq0.068$, so $\epsilon=0.01$ sits comfortably inside the budget, and $\epsilon=0.03$ slightly exceeds the first requirement but satisfies the second. 

\subsection{Spectrum selections}
We compare four combinations of spectra with the same set of free parameters: the $TT+TE+EE$ case uses the $T,E$ block of Eq.~\eqref{eq:covariance}; $EE+BB$ uses its diagonal $E,B$ block; the full configuration, labeled ``Total'' in the figures, uses all three fields; and $TE+BB$, which is a diagnostic test of the cross spectrum combined with $BB$. For Gaussian, parity-even fields, $\operatorname{Cov}(\widehat C_\ell^{TE},\widehat C_\ell^{BB})=0$, so the two Fisher contributions add, $\F^{TE+BB}=\F^{TE}+\F^{BB}$, with
\begin{equation}
\begin{split}
\mathcal F_{ij}^{TE}&=\sum_\ell
\frac{(2\ell+1)f_{\rm sky}(\ell)}{\Delta C_{\ell}^{TE}}
\frac{\partial C_\ell^{TE}}{\partial\theta_i}
\frac{\partial C_\ell^{TE}}{\partial\theta_j},\\
\Delta C_{\ell}^{TE}&=(C_\ell^{TT}+N_\ell^T)(C_\ell^{EE}+N_\ell^P)
+(C_\ell^{TE})^2,
\end{split}\label{eq:te}
\end{equation}
and
\begin{equation}
\mathcal F_{ij}^{BB}=\sum_\ell
\frac{(2\ell+1)f_{\rm sky}(\ell)}{2\left(C_\ell^{BB}+N_\ell^P\right)^2}
\frac{\partial C_\ell^{BB}}{\partial\theta_i}
\frac{\partial C_\ell^{BB}}{\partial\theta_j}.\label{eq:bb}
\end{equation}
In Eq.~\eqref{eq:te}, $TT$ and $EE$ enter only through the variance of $\widehat C_\ell^{TE}$, which is evaluated at the fiducial model.

Differences between the selections reflect both their direct tensor information and their ability to constrain the scalar parameters over which we marginalize. Their ranking depends on the adopted parameter set, on the noise, and on the absence of external priors.

Chiral gravitational waves, from parity-violating gravitational terms during inflation \cite{Lue1999,Saito2007} or from gauge fields (Sec.~\ref{sec:landscape}), produce nonzero $TB$ and $EB$ correlations, and so does cosmic birefringence, since it converts $E$ modes into $B$ modes \cite{Minami2020}. Our covariance sets these cross spectra to zero, so a chiral signal would carry information that we do not count, making our parity-even forecast conservative. Due to the $E$-mode leakage to $BB$, cosmic birefringence would also bias $r$, so a realistic analysis has to fit the rotation angle together with the tensor parameters. We leave both to future work.

\subsection{Delensing diagnostic}\label{sec:lensing}
Gravitational lensing of the CMB by large-scale structure converts $E$ modes into $B$ modes, producing a lensing $B$-mode component that adds to the primordial signal and, for a nearly scale-invariant primordial spectrum, contaminates the recombination bump. Throughout our baseline analysis, $C_\ell^{XY}$ are lensed spectra. To obtain a simple diagnostic of partial delensing, we decompose the $BB$ spectrum into an unlensed primary term, a lensing correction, and instrumental noise, and attenuate the lensing correction by a factor $\lambda$ through the template \cite{Campeti2019},
\begin{equation}
\widetilde C_\ell^{BB}(\lambda)
=C_{\ell,\rm unl}^{BB}
+\lambda\big(C_{\ell,\rm len}^{BB}-C_{\ell,\rm unl}^{BB}\big)
+N_\ell^{P},
\label{eq:delensing}
\end{equation}
while $TT$, $TE$ and $EE$ remain lensed. Here, $\lambda=1$ represents the undelensed baseline, while $\lambda=0.8$ corresponds to reducing the lensing contribution by $20\%$, a conservative benchmark representative of what can be achieved from internal CMB delensing alone, as we verify in Appendix~\ref{app:delensing} with an explicit reconstruction. We also consider $\lambda=0.6$, $0.4$, $0.2$, and $0.1$ for a qualitative representation of how delensing improves our constraints.

We also compute the residual left by specific tracers of the lensing potential: internal reconstruction, the cosmic infrared background (CIB), a Simons Observatory-like telescope, and their combinations; Appendix~\ref{app:delensing} gives the formalism. Internal reconstruction leaves $\lambda_{80}=0.81$, which motivates our benchmark, and the best combination reaches $0.43$.

We use this simplified prescription only to isolate the effect of reducing the lensing contribution: the reconstruction noise enters only through the correlation coefficient of the tracer, while the non-Gaussian covariance of the delensed spectrum and the $N^{(1)}$ reconstruction bias are not included.

\section{Results and discussion}\label{sec:results}

\subsection{Amplitude sensitivity and tilt marginalization}
\begin{figure*}[t]
\centering
\includegraphics[width=0.94\textwidth]{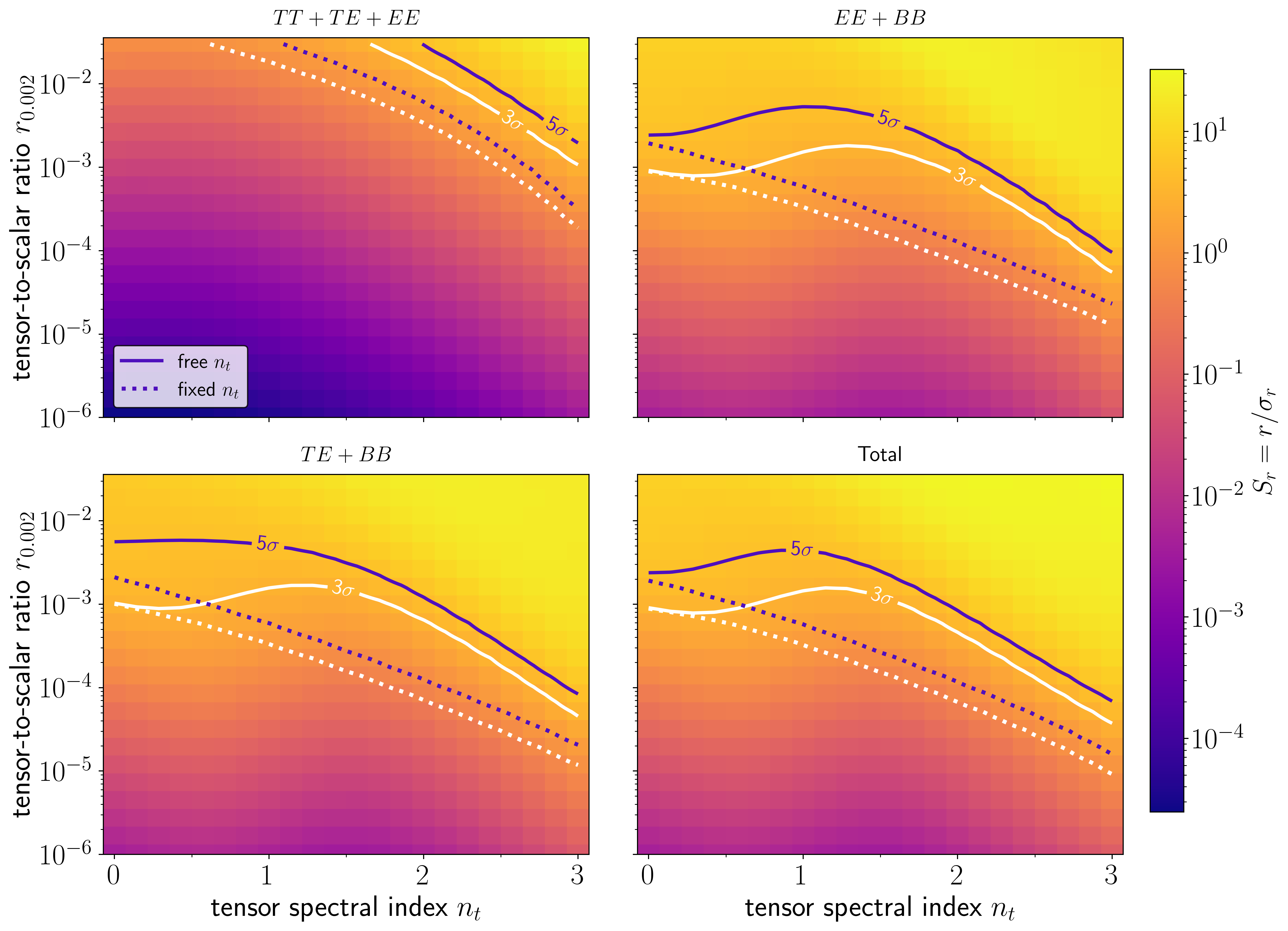}
\caption{Amplitude signal-to-noise ratio $\Sr=r/\sigma_r$ at $k_*=0.002\,\Mpc^{-1}$ for $TT+TE+EE$, $EE+BB$, $TE+BB$, and the full configuration (Total). The color scale shows the free-$n_t$ result, with the six $\LCDM$ parameters marginalized. White and purple curves mark $\Sr=3$ and $5$; solid and dotted lines correspond to free and fixed tilt, respectively.}\label{fig:amplitude}
\end{figure*}

Figure~\ref{fig:amplitude} shows the amplitude signal-to-noise ratio $\Sr$ for the four spectra selections, together with the contours $\Sr=3$ and $5$ for fixed and free tilt. At fixed $r_{0.002}$, increasing $n_t$ enhances the primordial power on scales smaller than the pivot ($k > k_*$) and shifts the information toward higher multipoles. The sensitivity therefore generally improves with both amplitude and tilt.

After marginalizing over $n_t$, however, the trend is not monotonic along every grid slice. The penalty is largest around $n_t\simeq1.5$, where $\sigma_r$ grows by a factor of about six at $r_{0.002}\simeq10^{-3}$, and decreases again for steeper spectra. The scalar parameters play a minor role: at $r_{0.002}=9.7\times10^{-4}$ and $n_t=1$, freeing only $n_t$ increases $\sigma_r$ by $261\%$, freeing only the six $\LCDM$ parameters by $15\%$, and freeing both by $274\%$. In the local Gaussian approximation, the penalty is set by the amplitude-tilt correlation $\rho_{rn_t}$, evaluated after marginalizing over the scalar parameters:
\begin{equation}
\frac{\sigma_{r,\rm marg}}{\sigma_{r,\text{fixed }n_t}}
=\frac{1}{\sqrt{1-\rho_{rn_t}^2}}.\label{eq:penalty}
\end{equation}

This behavior follows from the range of scales carrying the tensor information. For a nearly scale-invariant spectrum, the reionization and recombination $BB$ bumps provide information on both sides of the pivot $k_* = 0.002\,\Mpc^{-1}$, making the amplitude $r$ well-anchored and only weakly correlated with the tilt. At larger $n_t$, the signal shifts toward higher multipoles, so determining the normalization at the pivot requires a larger extrapolation and increases the correlation with $n_t$. For the steepest spectra, the tensor signal extends over a sufficiently broad multipole range that part of this degeneracy is broken.

\subsection{Sensitivity without \texorpdfstring{$B$}{B} modes}
The top-left panel of Fig.~\ref{fig:amplitude} quantifies what is achievable from temperature and $E$-mode polarization alone. The $TT+TE+EE$ combination reaches $\Sr=3-5$ only in the corner of large amplitude and extreme blue tilt; over most of the grid, and in particular near scale invariance, its amplitude sensitivity is poor. Marginalizing over $n_t$ pushes the threshold contours even further into that corner.

As expected, the tensor contribution to $TT$, $TE$, and $EE$ sits on top of the much larger scalar signal, whose cosmic variance acts as a fundamental noise floor. Unlike $BB$, which receives no scalar contribution at linear order, the temperature and $E$-mode spectra cannot separate a small tensor component from the dominant scalar one unless the former is unusually large. A blue tilt increases the relative tensor contribution at high multipoles, but temperature and $E$-mode spectra become competitive only for extreme signals that are already excluded by existing analyses \cite{Planck2018X,Galloni2023}. These spectra may nevertheless provide a useful cross-check independent of $B$-mode systematics such as $E\!\to\!B$ leakage, but they do not offer competitive sensitivity to small-amplitude tensors over most of the grid.

\subsection{Polarization subsets}\label{sec:subsets}
\begin{figure*}[t]
\centering
\includegraphics[width=\textwidth]{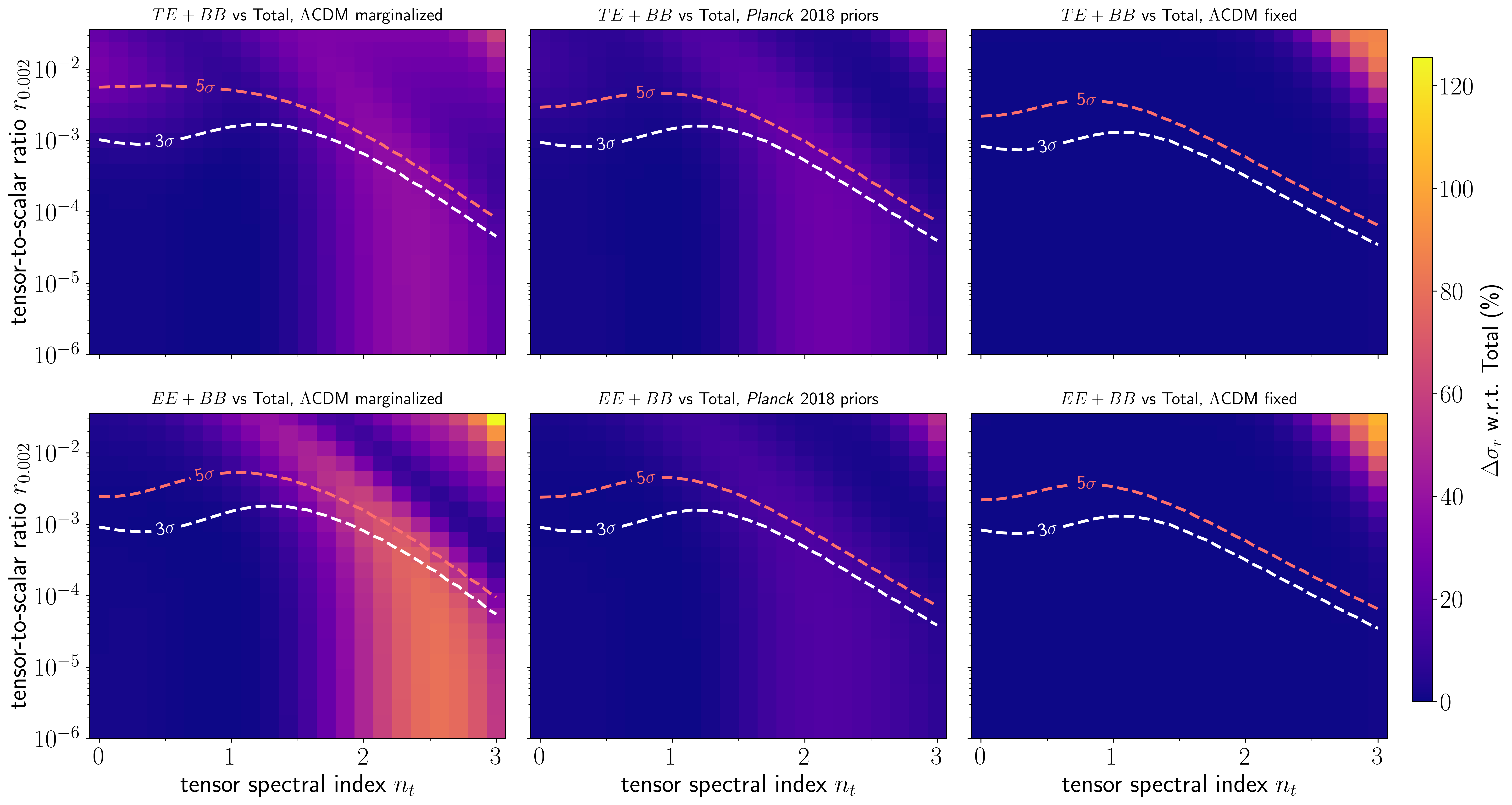}
\caption{Fractional increase of the marginalized amplitude error, $100(\sigma_r^{\rm subset}/\sigma_r^{\rm Total}-1)$, for $TE+BB$ (top) and $EE+BB$ (bottom), with $n_t$ free and the scalar parameters marginalized (left), constrained by \textit{Planck} 2018 priors (center), or fixed (right). White and orange dashed curves mark $\Sr=3$ and $5$ for the respective subset.}\label{fig:subsets}
\end{figure*}

The three combinations that include $B$-mode information constrain a much larger region of the plane than the $TT{+}TE{+}EE$ block, and their contours lie close to one another. The direct sensitivity to $r$ comes primarily from $BB$, while scalar spectra help to constrain the scalar parameters and break degeneracies. Figure~\ref{fig:subsets} compares the two polarization subsets with the full configuration for the three treatments of the scalar parameters. With the scalar parameters marginalized, the two subsets show similar patterns of information loss: compared with Total, $\sigma_r$ is larger by $16.5\%$ for $TE+BB$ and by $13.8\%$ for $EE+BB$ at the median grid point, and $EE+BB$ inflates it by up to a factor of 2.26 in the corner at $r_{0.002}=3\times10^{-2}$ and $n_t=3$.  
\textit{Planck} priors reduce the median increases to $7.2\%$ and $4.4\%$, and $EE+BB$ becomes slightly better than $TE+BB$. With the scalar parameters fixed, both median increases drop to $0.1\%$, and $\sigma_r$ from $EE+BB$ varies between $-4\%$ and $+9\%$ with respect to the $TE+BB$ configuration. The only residual loss with respect to Total appears for both subsets at $r_{0.002}\gtrsim10^{-2}$ and $n_t\gtrsim2.5$, where the tensor contribution to $TT$ is no longer negligible.

The difference between these subsets therefore comes from the different constraining power that the two spectra have on the scalar cosmology, not from direct tensor information in $TE$, and it does not suggest a general advantage of $TE$ over $EE$ for tensor detection.

\subsection{Effect of delensing}
\begin{figure*}[t]
\centering
\includegraphics[width=\textwidth]{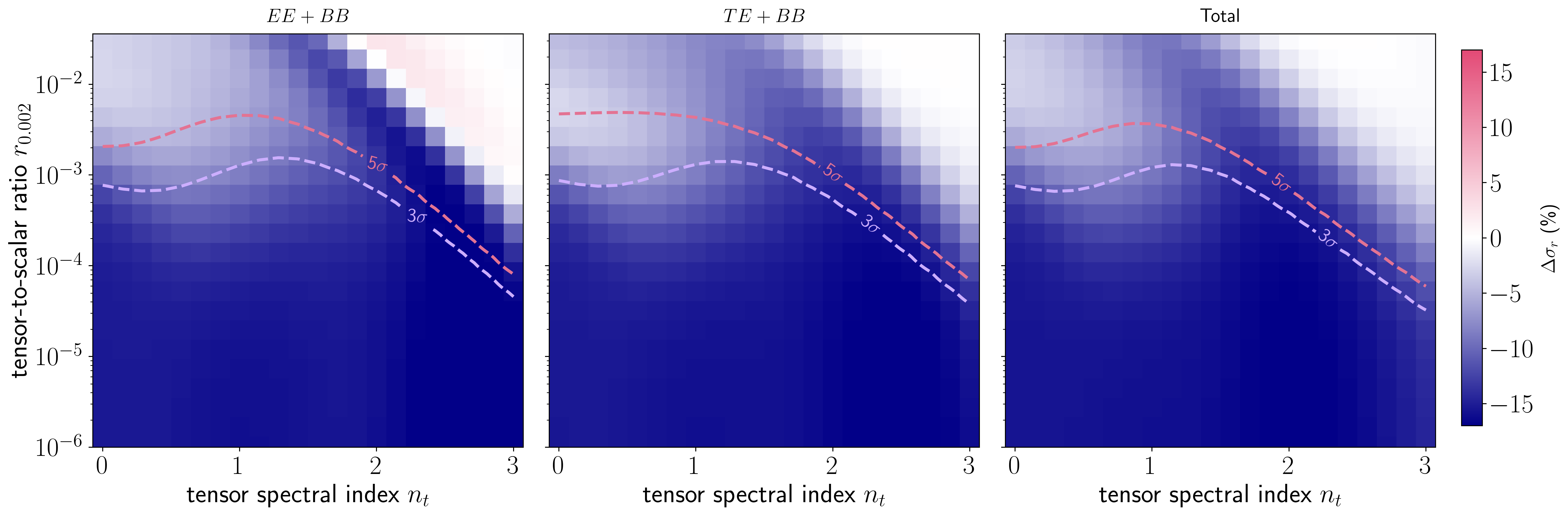}
\caption{Fractional change of the marginalized amplitude error, $100[\sigma_r(\lambda=0.8)/\sigma_r(\lambda=1)-1]$, when the lensing $B$-mode power is reduced by $20\%$. Negative values indicate smaller errors. Purple and pink dashed curves mark delensed $\Sr=3$ and $5$.}\label{fig:delensing}
\end{figure*}

For configurations containing $BB$, a $20\%$ reduction in the lensing $B$-mode power improves the marginalized constraint on $r$ by at most $17\%$ for Total and $18\%$ for the two subsets, each compared with the same combination without delensing (Fig.~\ref{fig:delensing}). The improvement is largest at low tensor amplitudes, where lensing $B$-modes contribute substantially to the variance, and becomes smaller as the primordial signal dominates the lensing contribution.

Including delensing also in $TT$, $TE$ and $EE$ changes these improvements by at most 2.2 percentage points with respect to the $BB$-only case, but in the extreme upper-right corner of the grid it makes $\sigma_r$ up to $2\%$ larger than without delensing, because it removes scalar information from those spectra without adding the information from the lensing reconstruction. With the reduction applied to $BB$ only, the increase disappears for Total (at most $0.1\%$) and is at most $2.4\%$ for $EE+BB$, where the lensing $B$ modes help to constrain the scalar parameters, and we checked that adding the lensing reconstruction as an additional observable restores this loss.

Because it is limited to a $20\%$ reduction, the modest gain found here does not mean that delensing is unimportant. Stronger delensing, using external tracers or high-resolution CMB observations, can yield larger improvements \cite{SimonsObs2019,Namikawa2016}. For the Total configuration, compared with no delensing, $\sigma_r$ decreases at $r_{0.002}=9.7\times10^{-4}$ and $n_t=0$ by $10\%$, $31\%$ and $45\%$ for $\lambda=0.8$, $0.4$ and $0.1$, and at $r_{0.002}=1.4\times10^{-4}$ and $n_t=2$ by $16\%$, $49\%$ and $71\%$. With the tracers of Appendix~\ref{app:delensing}, the median improvement of $\sigma_r$ over the grid points with $\Sr>3$ ranges from $6\%$ for internal reconstruction to $18\%$ for the Simons Observatory-like estimator combined with the CIB.

\subsection{Joint amplitude-tilt sensitivity}\label{sec:joint}
\begin{figure*}[t]
\centering
\includegraphics[width=0.94\textwidth]{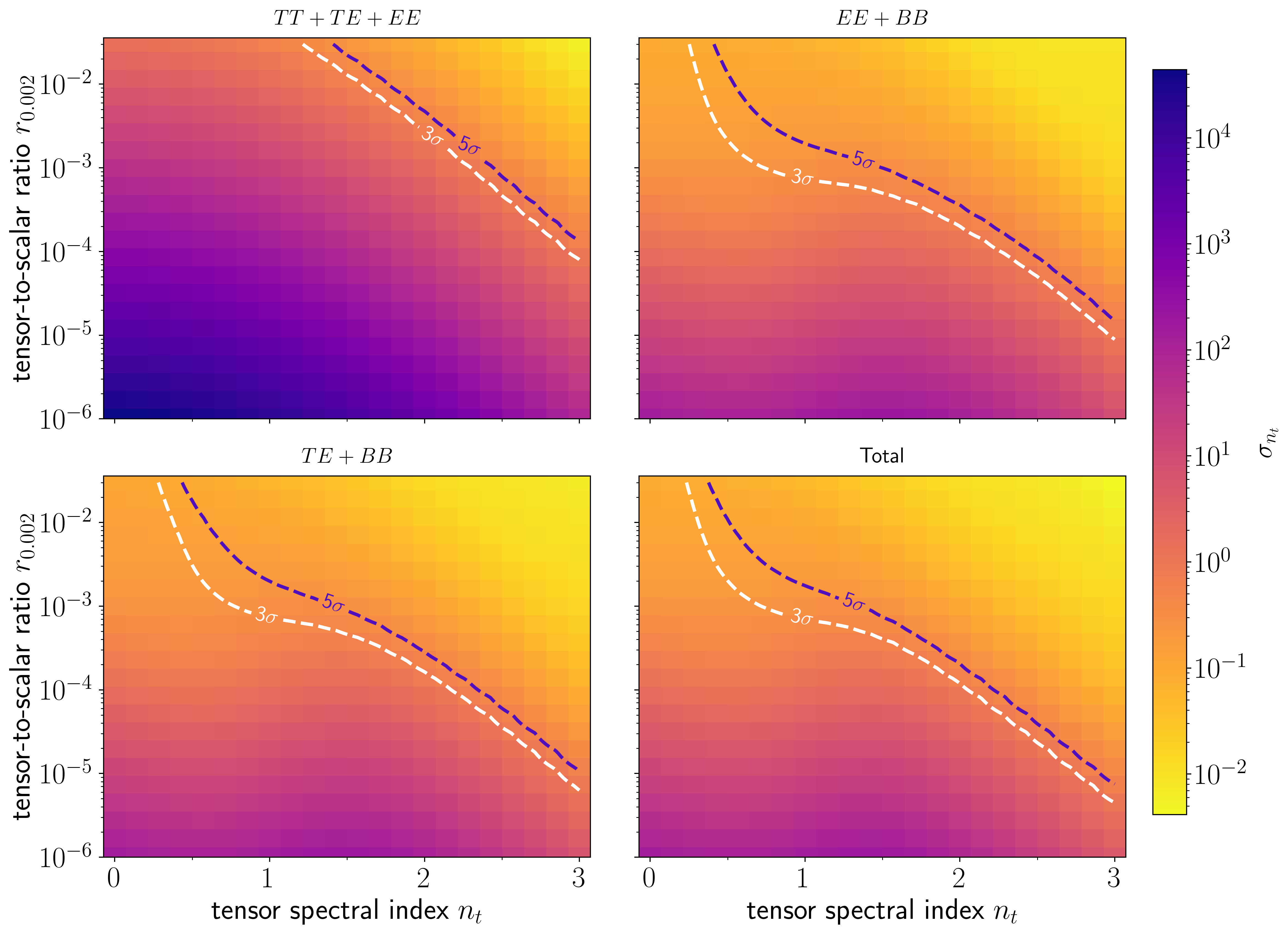}
\caption{Marginalized uncertainty $\sigma_{n_t}$ for the four selections, with $r$ and the scalar parameters free. White and purple dashed curves mark $\Snt=3$ and $5$; the $\Snt=2$ contour used in Eq.~\eqref{eq:characterization} is shown in Fig.~\ref{fig:joint}.}\label{fig:tilt}
\end{figure*}

While detecting the tensor amplitude confirms the presence of a gravitational wave background, discriminating among the scenarios of Sec.~\ref{sec:landscape}, and separating them from canonical inflation, requires measuring the tilt. Figure~\ref{fig:tilt} shows $\sigma_{n_t}$ after marginalizing over $r$ and the scalar parameters. At fixed tilt, the uncertainty grows as the amplitude decreases toward $r=0$, where the slope becomes undefined. At fixed amplitude, a bluer tilt improves the slope measurement by increasing the tensor signal over a broader range of scales. The ranking of the four selections follows that of the amplitude analysis: $TT+TE+EE$ is informative only in the extreme part of the grid, $BB$ greatly extends the sensitive region, and, with the scalar parameters marginalized, $TE+BB$ characterizes a slightly larger part of the grid than $EE+BB$ ($34.1\%$ against $33.5\%$, Table~\ref{tab:subsets}), for the reasons discussed in Sec.~\ref{sec:subsets}.

To identify fiducial models for which both the normalization and the sign of the slope are locally measurable, we define the characterization region
\begin{equation}
\mathcal R_{\rm char}=\{(n_t,r):\ \Sr>3\ \text{and}\ \Snt>2\}.\label{eq:characterization}
\end{equation}
The first condition requires that the tensor amplitude at $k_*=0.002\,\Mpc^{-1}$ is measured with sufficient significance that the characterization region is safely away from the unphysical $r<0$ boundary and the $r=0$ statistical pathologies, ensuring that the local Gaussian approximation remains robust. The second requires the positive tensor tilt to be separated from zero by at least two marginalized standard deviations. We use these two thresholds as practical criteria for model characterization rather than as a two-dimensional confidence contour.

Within the Gaussian approximation, $\Snt$ also gives the local significance for rejecting the $n_t=0$ hypothesis, as described in Appendix~\ref{app:statistics}. There, we also discuss how our characterization requirements differ from the formal definition of a joint two-dimensional confidence ellipse. Although canonical inflation predicts a slight red tilt ($n_t = -r/8$), this deviation lies well below our experimental resolution. Adopting $n_t = 0$ is practically identical for our purposes and acts as our null hypothesis.

\begin{figure*}[t]
\centering
\includegraphics[width=0.94\textwidth]{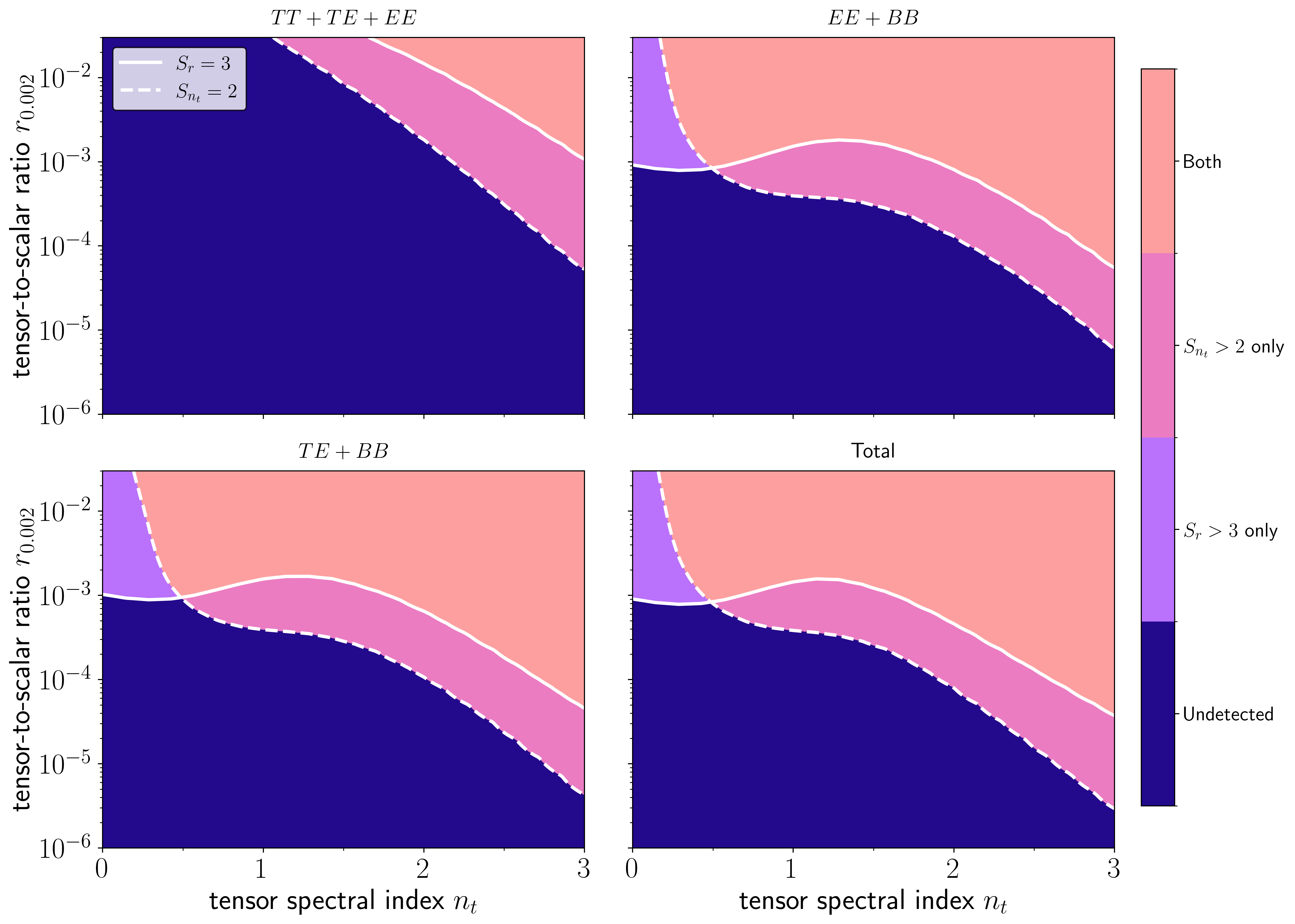}
\caption{Selection map defined by Eq.~\eqref{eq:characterization}. Colors indicate the regions where neither criterion is satisfied (labeled ``Undetected''), only $\Snt>2$, only $\Sr>3$, or both. Solid and dashed white curves denote the $\Sr=3$ and $\Snt=2$ thresholds, respectively, with continuous boundaries obtained by linear interpolation across the grid points.}\label{fig:joint}
\end{figure*}

Figure~\ref{fig:joint} shows the resulting selection map, where we interpolate linearly across the grid to determine the continuous boundaries of the significance regions. In the full configuration, joint detection and characterization ($S_r > 3$ and $S_{n_t} > 2$) first opens up at $n_t \simeq 0.17$ for amplitudes near the upper edge of our grid, $r_{0.002} \simeq 2.6\times 10^{-2}$. As the tilt grows steeper, lower tensor amplitudes are allowed into the characterization region. About $92\%$ of the area that passes the amplitude cut also passes the tilt cut, so that about one third ($36.4\%$) of the grid satisfies both criteria. 

\subsection{How robust is the characterization region?}\label{sec:robust}
\begin{figure*}[t]
\centering
\includegraphics[width=\textwidth]{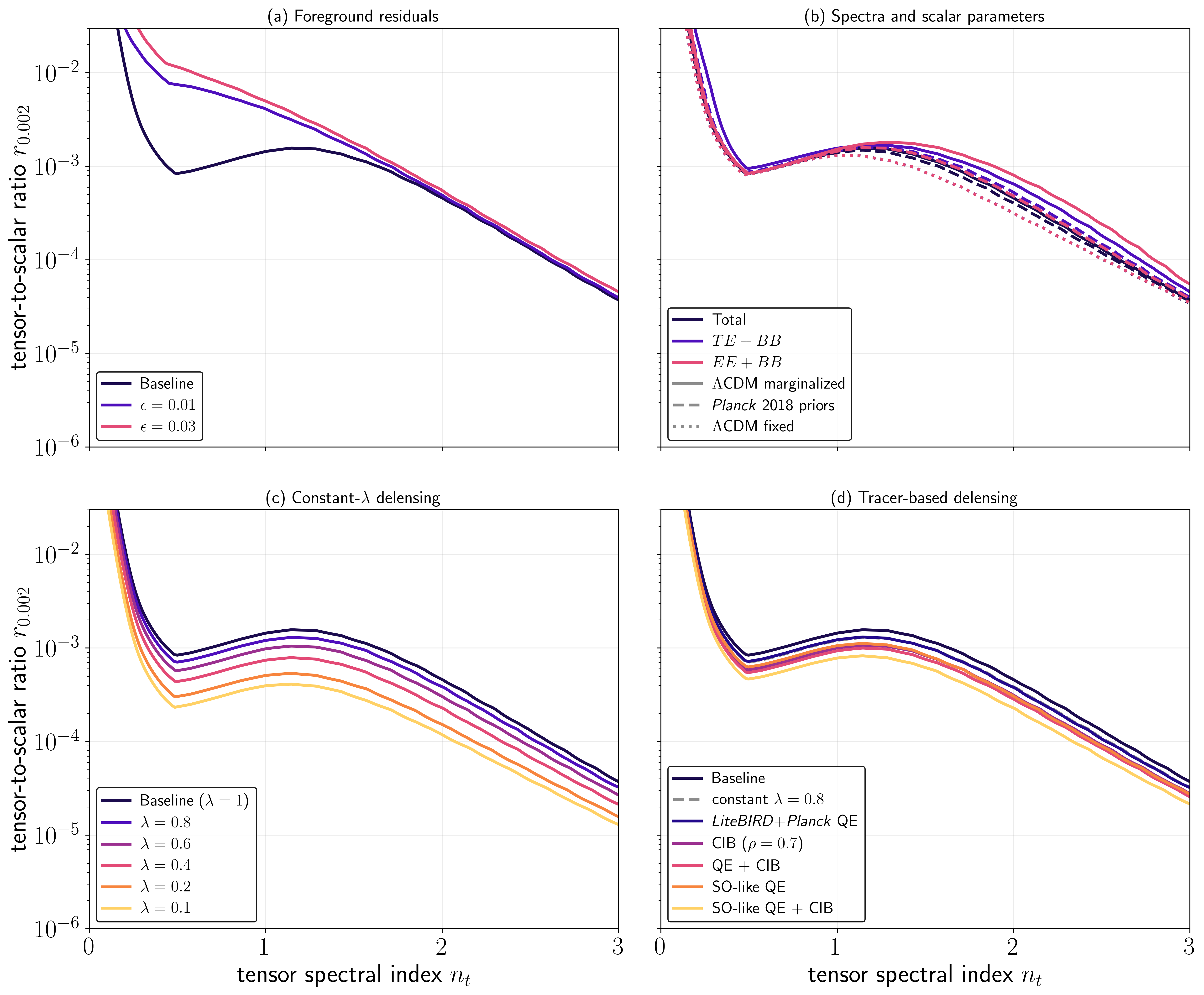}
\caption{Boundary of the characterization region ($\Sr>3$ and $\Snt>2$; the region lies above each curve). Baseline: Total, no foreground residual, $\LCDM$ marginalized, no delensing. (a) Foreground residual with $\epsilon=0.01$ and $0.03$ (Sec.~\ref{sec:fg}). (b) Spectrum combinations (colors) and treatment of the scalar parameters (line styles). (c) Constant residual lensing fraction $\lambda$ applied to $BB$. (d) Delensing with specific tracers, compared with the constant $\lambda=0.8$.}\label{fig:curves}
\end{figure*}

We now study how the characterization region changes as we add foreground residuals, select subsets of spectra, change the treatment of the scalar parameters and delens the $B$ modes. The reference is the baseline of Fig.~\ref{fig:joint}: Total, with no foreground residual, the six $\LCDM$ parameters marginalized and no delensing. The results are shown in Fig.~\ref{fig:curves}, and Table~\ref{tab:robust} lists the characterized fraction and the point where the region opens.

Foregrounds remove about a sixth of the region, and the loss is localized below $n_t=1.5$, where the characterized share of the plane falls from $25.8\%$ to $15.1\%$ for $\epsilon=0.01$ and to $12.8\%$ for $\epsilon=0.03$, while above $n_t=2$ it barely moves, from $52.8\%$ to $52.2\%$ and $51.1\%$. As anticipated in Sec.~\ref{sec:method}, foregrounds affect the region where the tensor signal is still significant at low multipoles, and a bluer spectrum moves the signal away from them. The onset shifts from $n_t\simeq0.17$ to $0.21$ and $0.29$ for $\epsilon=0.01$ and 0.03 respectively.

The treatment of the scalar parameters affects the characterization much less: for the Total configuration, the region occupies between $36.4\%$ and $38.4\%$ of the whole grid (Table~\ref{tab:subsets}), and the curves in panel (b) differ mainly at $n_t\gtrsim1.5$, where the tilt-amplitude degeneracy is strongest.

\begin{table}[t]
\caption{Characterized fraction of the grid ($\Sr>3$ and $\Snt>2$, \%) for each combination and treatment of the $\LCDM$ parameters.}\label{tab:subsets}
\begin{tabular}{lccc}
\hline\hline
 & marginalized & \textit{Planck} priors & fixed\\
\hline
$TE+BB$ & 34.1 & 35.6 & 38.4\\
$EE+BB$ & 33.5 & 36.2 & 38.4\\
Total & 36.4 & 37.0 & 38.4\\
\hline\hline
\end{tabular}
\end{table}

Reducing the lensing $B$-mode power shifts the whole boundary to lower amplitudes, and the characterized fraction grows coherently with the reduction: from $36.4\%$ without delensing to $38.0\%$, $40.0\%$, $42.6\%$, $46.0\%$ and $48.3\%$ for $\lambda=0.8$, $0.6$, $0.4$, $0.2$ and $0.1$, while the onset moves from $n_t\simeq0.17$ to $0.12$. Realistic tracers display a similar behavior: internal reconstruction ($\lambda_{\ell=80}\equiv \lambda_{80}=0.81$) gives $38.1\%$ and reproduces the constant $\lambda=0.8$ almost exactly, the CIB and the SO-like estimator give $39.8\%$ and $39.6\%$ respectively, and their combination ($\lambda_{80}=0.43$) reaches $42.3\%$, close to a constant $\lambda=0.4$.

\subsection{Can a tilt of \texorpdfstring{$n_t\simeq0.034$}{nt=0.034} be measured?}\label{sec:stringgas}
\begin{figure*}[t]
\centering
\includegraphics[width=0.94\textwidth]{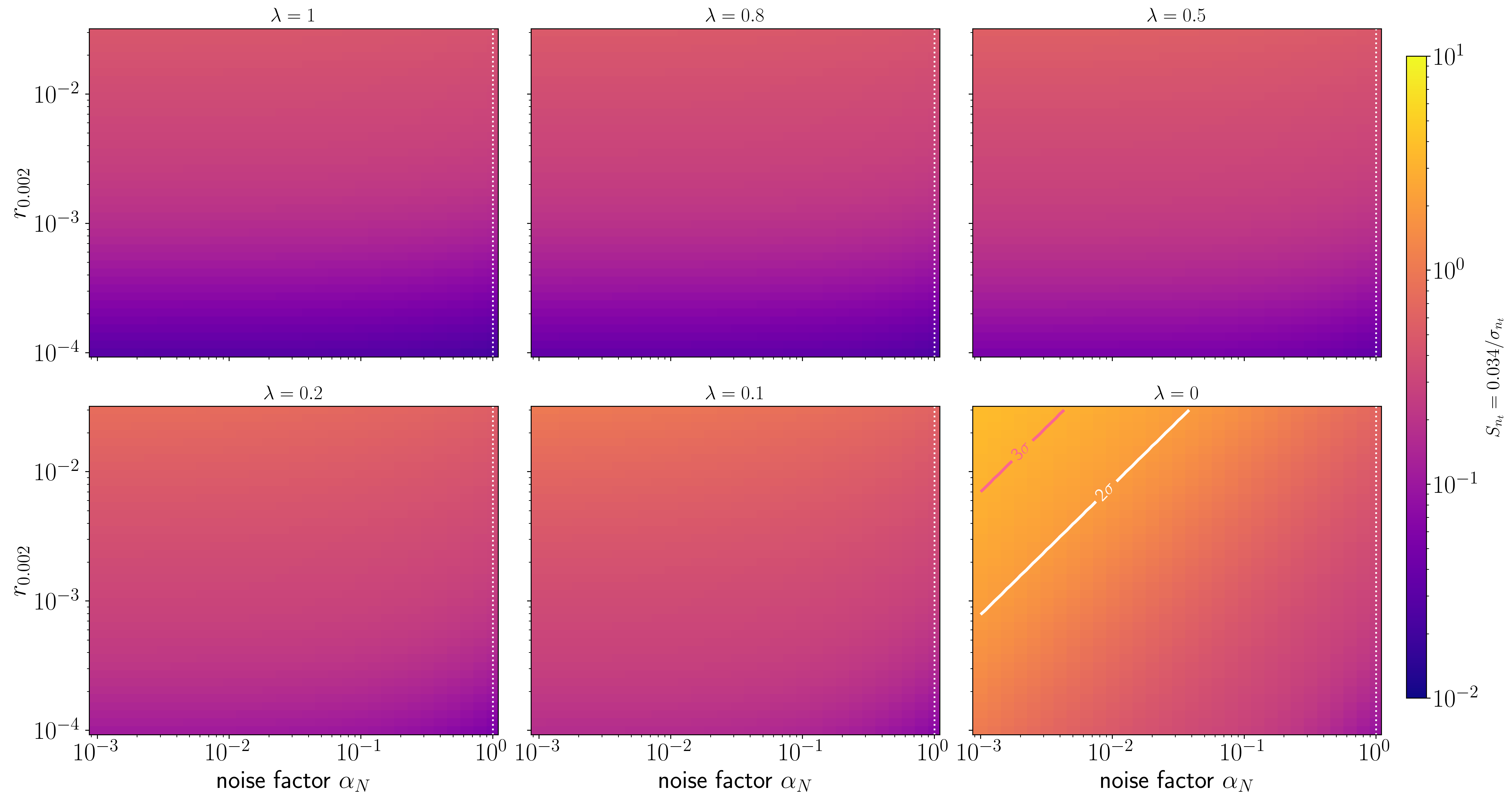}
\caption{Significance $\Snt=0.034/\sigma_{n_t}$ (Total, $\LCDM$ marginalized) as a function of the noise-reducing factor $\alpha_N$ and of $r_{0.002}$, for six values of the residual lensing fraction $\lambda$; contours mark $\Snt=2$ and 3.}\label{fig:stringgas}
\end{figure*}

The string-gas tilt of Eq.~\eqref{eq:stringgas} lies below our first positive grid value, so we scan $\Snt$ for $n_t=0.034$ over the amplitude range $[10^{-4},0.03]$, scaling the polarization noise power of \textit{LiteBIRD} ($2\leq\ell\leq600$) by a factor $\alpha_N$ and the lensing $B$-mode power by $\lambda$ (Fig.~\ref{fig:stringgas}). With our noise and no delensing, $\Snt$ is $0.40$ at $r_{0.002}=2.6\times10^{-2}$ ($\sigma_{n_t}\simeq0.085$), which is about the largest amplitude allowed for this tilt by the bound $r_{0.01}<0.029$ (Sec.~\ref{sec:constraints}), while the characterization $\Snt=2$ needs $\sigma_{n_t}\simeq0.017$. Pushing towards cosmic variance, $\alpha_N\rightarrow0$, does not guarantee a detection: whenever $\lambda\geq0.1$, $\Snt$ stays below 1 for every noise level and every $r_{0.002}\leq0.03$, because lensing sample variance takes over. Only full delensing, $\lambda=0$, allows characterization: $\Snt=2$ needs $\alpha_N\simeq0.03$ at $r_{0.002}\simeq2.5\times10^{-2}$, and $\alpha_N=10^{-3}$ brings the threshold down to $r_{0.002}\simeq8\times10^{-4}$. Fixing the scalar parameters does not change the picture: for $\lambda=0.1$ the best value of $\Snt$ rises from $0.99$ to $1.27$, still much below 2. A significant measurement of the string-gas tilt would therefore need lensing removed entirely and polarization sensitivity about five times better than that of \textit{LiteBIRD}, with an amplitude close to the present bound.

\subsection{Present constraints on the plane}\label{sec:constraints}
We now compare our characterization region with present constraints to see which part of the plane survives the selection. Table~\ref{tab:bounds} summarizes the bounds that apply to our plane, with their pivots and assumptions. Equation~\eqref{eq:pivot} maps a bound on $r(k_c)$ onto $k_*$, and for a blue tilt the mapping is quite steep, as already mentioned in Sec. \ref{sec:spectrum}. We consider only analyses in which $n_t$ is varied.

The LIGO-Virgo-KAGRA (LVK) limit is quoted for a power law in frequency, $\Omega_{\rm GW}(f)=\Omega_{\rm ref}(f/25\,{\rm Hz})^{\alpha_{\rm GW}}$, with separate upper limits on $\Omega_{\rm ref}$ for $\alpha_{\rm GW}=0$, $2/3$ and $3$ \cite{LVK2021}, where $\alpha_{\rm GW}$ is the spectral slope of the energy density. To check which regions survive the bound, we compare $\mathcal W(x)\Pt$ at $25\,$Hz with the limit interpolated in $\alpha_{\rm GW}$. The $\Delta N_{\rm eff}$ bound uses Eq.~\eqref{eq:bbn} with $\Delta N_{\rm eff}<0.3$ \cite{Planck2018VI}, integrated from $10^{-10}\,$Hz, roughly the frequency of the modes inside the horizon at nucleosynthesis, to $10^{8}\,$Hz, roughly the mode that left the horizon at the end of inflation.

\begin{table*}[t]
\caption{Present bounds used in Fig.~\ref{fig:constraints} (95\%).}\label{tab:bounds}
\begin{tabular}{p{5.6cm}p{9.6cm}l}
\hline\hline
Bound & Data and assumptions & Ref.\\
\hline
$r_{0.01}<0.029$, $-1.21<n_t<3.54$ & \textit{Planck} 2018 with the PR4 low-$\ell$ polarization likelihood (PL21) and BICEP/Keck 2018, free $n_t$ & \cite{Galloni2023}\\
$r_{0.01}<0.028$, $-1.37<n_t<0.42$ & as above, with the LVK O3 limit, free $n_t$ & \cite{Galloni2023}\\
$r_{0.002}<0.05$, $-0.6<n_t<2.6$ & \textit{Planck} 2018 with BICEP/Keck 2015, free $n_t$ & \cite{Planck2018X}\\
$\Omega_{\rm GW}(25\,{\rm Hz})<5.8\times10^{-9}$, $3.4\times10^{-9}$, $3.9\times10^{-10}$ & LVK O3 isotropic background, $\alpha_{\rm GW}=0$, $2/3$, $3$, log-uniform prior on $\Omega_{\rm ref}$, $H_0=67.9$ & \cite{LVK2021}\\
$\Delta N_{\rm eff}<0.30$ & \textit{Planck} 2018 $TT,TE,EE$+low$T$+low$E$+lensing with BAO & \cite{Planck2018VI}\\
\hline\hline
\end{tabular}
\end{table*}

\begin{figure}[t]
\centering
\includegraphics[width=\columnwidth]{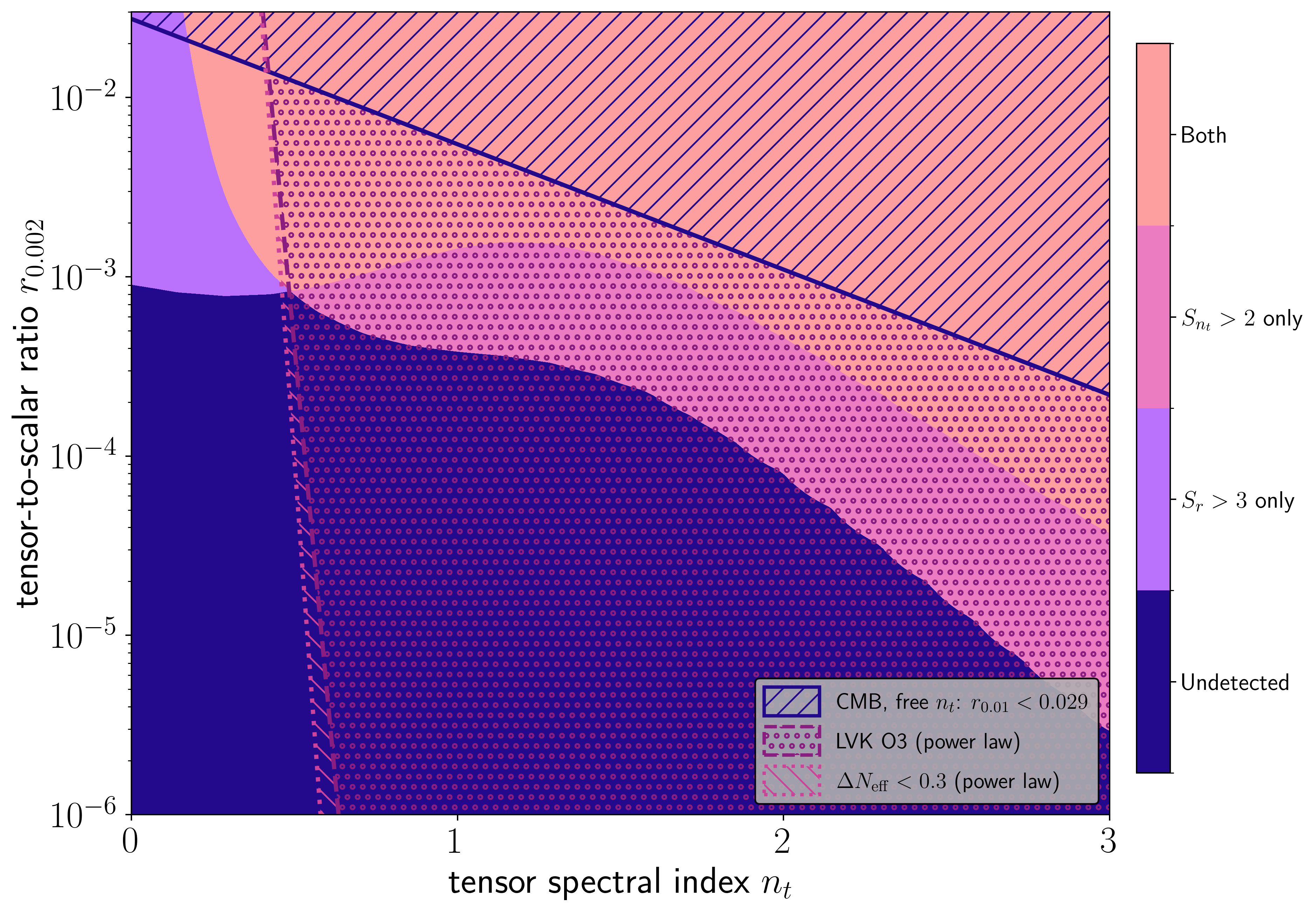}
\caption{Selection map of Fig.~\ref{fig:joint} (Total) with present bounds: CMB-only free tilt (solid line, diagonal hatching), LVK O3 (dashed line with circles) and $\Delta N_{\rm eff}$ (dotted line, reverse diagonal hatching), the last two assuming a power law at all frequencies. The allowed side is always to the left.}\label{fig:constraints}
\end{figure}

Figure~\ref{fig:constraints} overlays these bounds on the characterization region, with each excluded region hatched. For the CMB-only comparison we use the more stringent in $r$ of the two free-tilt bounds of Table~\ref{tab:bounds}, $r_{0.01}<0.029$. It encloses all the tilts used in our analysis, allowing about three quarters of the grid and $33\%$ of the characterization region. Adding LVK leaves $4.3\%$ of the characterization region, and including $\Delta N_{\rm eff}$ leaves $4.0\%$. The surviving part is a narrow portion at $0.2\lesssim n_t\lesssim0.45$ and $10^{-3}\lesssim r_{0.002}\lesssim2\times10^{-2}$.

\begin{figure}[t]
\centering
\includegraphics[width=\columnwidth]{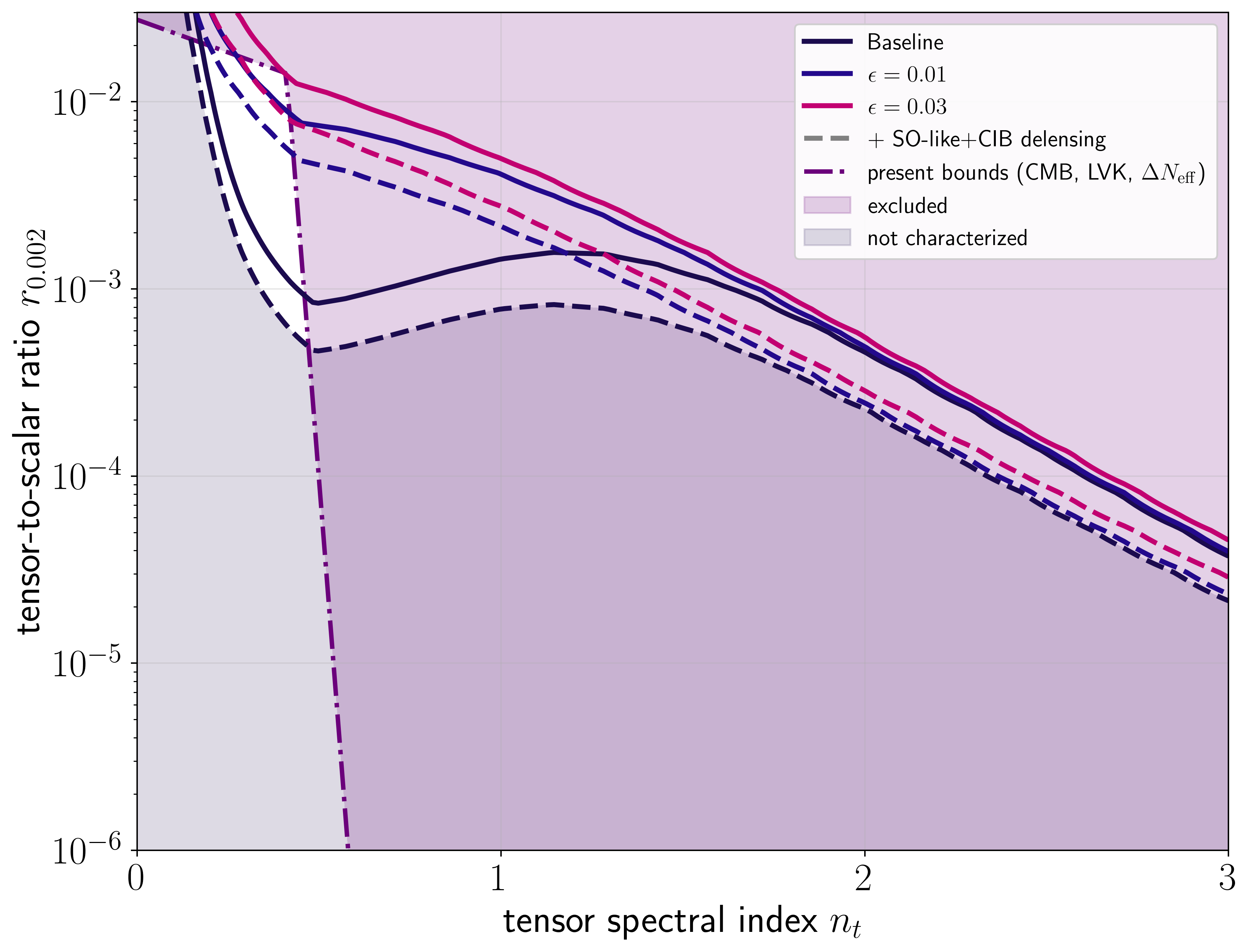}
\caption{Characterization boundary (Total, $\LCDM$ marginalized) for no foreground residual and $\epsilon=0.01$ and $0.03$, without delensing (solid) and with SO+CIB delensing (dashed, Appendix~\ref{app:delensing}). The dash-dotted line is the most stringent present bound at each $r_{0.002}$; the shaded area to its right is excluded. The grey zone is not characterized in any case.}\label{fig:fgdelens}
\end{figure}

Foreground residuals shrink the region even further: as they raise the boundary of the characterization region at low $n_t$, which is the only range the power-law bounds leave open, only $0.4\%$ of the characterization region survives all bounds for $\epsilon=0.01$, and for $\epsilon=0.03$ no region is allowed. As a fraction of the whole grid, the surviving area falls from $1.4\%$ for the baseline, to $0.1\%$ for $\epsilon=0.01$ and then to zero for $\epsilon=0.03$. Delensing instead recovers part of the characterization region: with the SO-like reconstruction combined with the CIB (Appendix~\ref{app:delensing}), the characterized fraction becomes $36.2\%$ for $\epsilon=0.01$ and $34.2\%$ for $\epsilon=0.03$ (Table~\ref{tab:robust}), and the area allowed by all bounds grows to $0.40\%$ and $0.14\%$ of the grid (Fig.~\ref{fig:fgdelens}). Without a foreground residual, the same tracer combination SO+CIB allows $2.0\%$ of the total grid to survive the bounds. For $\epsilon=0.03$ no part of the plane is both characterized and allowed unless the $B$ modes are delensed, and for $\epsilon=0.01$ delensing more than triples the surviving area. The direct CMB bound alone is less severe: with foreground residuals of $\epsilon=0.01$ and $0.03$ and the SO+CIB delensing, it keeps $33.0\%$ and $29.2\%$ of the respective characterization regions.

The interferometer and $\Delta N_{\rm eff}$ bounds assume that the same power law holds from $10^{-18}\,$Hz up to $25\,$Hz or beyond. A running, a break, a cutoff or a non-standard thermal history (Sec.~\ref{sec:plane}) would relax them and open a much larger part of the plane on CMB scales. Only the region excluded by CMB data alone is independent of that extrapolation.

\begin{table*}[t]
\caption{Characterized fraction of the grid and onset of the characterization region (smallest $n_t$ and the corresponding $r_{0.002}$) for Total with $\LCDM$ marginalized, and percentage of the grid that is characterized and allowed by all present bounds of Sec.~\ref{sec:constraints} (CMB, LVK O3 and $\Delta N_{\rm eff}$).}\label{tab:robust}
\begin{tabular}{lcccc}
\hline\hline
Case & characterized (\% of grid) & onset $n_t$ & $r_{0.002}$ at onset & allowed (\% of grid)\\
\hline
Baseline & 36.4 & 0.17 & $2.6\times10^{-2}$ & 1.44\\
$\epsilon=0.01$ & 30.7 & 0.21 & $2.8\times10^{-2}$ & 0.12\\
$\epsilon=0.03$ & 29.0 & 0.29 & $2.8\times10^{-2}$ & 0\\
\textit{LiteBIRD} 2026 & 36.1 & 0.17 & $2.7\times10^{-2}$ & 1.41\\
$\lambda=0.8$ & 38.0 & 0.17 & $2.2\times10^{-2}$ & 1.60\\
$\lambda=0.6$ & 40.0 & 0.15 & $2.6\times10^{-2}$ & 1.80\\
$\lambda=0.4$ & 42.6 & 0.14 & $2.8\times10^{-2}$ & 2.08\\
$\lambda=0.2$ & 46.0 & 0.12 & $2.8\times10^{-2}$ & 2.51\\
$\lambda=0.1$ & 48.3 & 0.12 & $2.1\times10^{-2}$ & 2.82\\
Internal QE & 38.1 & 0.17 & $2.2\times10^{-2}$ & 1.59\\
CIB ($\rho=0.7$) & 39.8 & 0.15 & $2.7\times10^{-2}$ & 1.79\\
Internal QE + CIB & 40.5 & 0.15 & $2.4\times10^{-2}$ & 1.85\\
SO-like QE & 39.6 & 0.15 & $2.8\times10^{-2}$ & 1.72\\
SO-like QE + CIB & 42.3 & 0.14 & $3.0\times10^{-2}$ & 2.02\\
$\epsilon=0.01$, SO-like QE + CIB & 36.2 & 0.17 & $3.0\times10^{-2}$ & 0.40\\
$\epsilon=0.03$, SO-like QE + CIB & 34.2 & 0.21 & $2.8\times10^{-2}$ & 0.14\\
\hline\hline
\end{tabular}
\end{table*}

\subsection{Implications for the model landscape}\label{sec:implications}
\textit{Moderately blue models.} Models with an appreciable CMB-scale amplitude and positive tilt are the most promising targets for CMB measurements of a blue tensor spectrum. If the spectrum keeps the same power law up to interferometer scales, present data restrict the characterizable ones to $n_t\lesssim0.45$. The representative string-gas value $n_t\simeq0.034$ lies below the first positive grid value, and Sec.~\ref{sec:stringgas} shows that measuring it would require near-complete delensing and much lower noise.

\textit{Strongly blue models.} Representative pre-big-bang and ekpyrotic spectra have large-scale amplitudes far below the range explored here. Despite their steep tilts, detectability in the CMB is controlled by the tensor normalization at cosmological scales. High-frequency observations may provide stronger constraints, but this requires a specification of the spectral shape, thermal history, and integrated energy-density bound.

\textit{Sourced and transfer-induced branches.} Smooth portions of sourced spectra can be approximated locally by a power law, whereas sharp features, strong running, chirality, and significant non-Gaussianity require full templates. In particular, a kination-induced high-frequency enhancement is instead a transfer effect and does not correspond to a positive primordial CMB tilt.

\section{Conclusions}\label{sec:conclusions}

We reviewed representative mechanisms for blue tensor spectra and calculated the CMB sensitivity to $r$ and $n_t$ in an idealized \textit{LiteBIRD}+\textit{Planck} configuration. The forecast varies the six scalar parameters together with $r$ and $n_t$ and compares four combinations of CMB spectra over a $22\times22$ grid, testing the results against a foreground residual, priors on the scalar parameters, delensing, and present observational bounds.

The tensor amplitude is generally easier to measure for larger $r$ and bluer spectra, although letting the tilt vary substantially weakens the constraint on the normalization at the large-scale pivot; marginalizing over the scalar parameters adds a much smaller penalty in comparison. The $TT+TE+EE$ configuration reaches $\Sr=3-5$ only for extreme signals, whereas the selections containing $BB$ probe a much larger part of the grid. The comparison of polarization subsets illustrates the importance of scalar-parameter degeneracies in determining the marginalized tensor constraint. With the scalar parameters marginalized, $TE+BB$ remains closer to the full result than $EE+BB$ and characterizes a larger fraction of the grid ($34.1\%$ against $33.5\%$ respectively), while for $EE+BB$ the amplitude error increases by up to a factor 2.26. With \textit{Planck} priors or fixed scalar parameters, instead, the two subsets become equivalent. Partial delensing of the $B$ modes with $\lambda=0.80$ provides a modest improvement, reaching roughly $17\%$ in $\sigma_r$ in the lensing-dominated regime, and between $45\%$ and $71\%$ for a residual lensing fraction of $0.1$.

Requiring $\Sr>3$ and $\Snt>2$ selects about one third of the whole grid. The characterization region opens up at $n_t \simeq 0.17$ for $r_{0.002} \simeq 2.6\times 10^{-2}$ and broadens toward lower amplitudes as the tilt increases. A foreground residual at the level of the \textit{LiteBIRD} error budget reduces the characterized fraction from $36\%$ to between $29\%$ and $31\%$, with the loss concentrated at $n_t<1.5$, while delensing with external tracers increases it up to $42\%$. Measuring the string-gas tilt $n_t\simeq0.034$ would require nearly complete delensing and a polarization noise power about thirty times lower than \textit{LiteBIRD}. Present CMB bounds leave $33\%$ of the characterization region, and only about $4\%$ of it survives if the power law is extended up to the LVK band and over the frequencies that contribute to $\Delta N_{\rm eff}$, a cut that a running, a break or a cutoff would remove. With a foreground residual this surviving fraction drops to $0.4\%$ for $\epsilon=0.01$ and vanishes for $\epsilon=0.03$; delensing with an external tracer, such as a high-resolution lensing reconstruction combined with the CIB, is then needed to keep a region that is both characterized and allowed. In this forecast, mildly blue models with a detectable amplitude stand out as the primary targets for future CMB experiments. This result concerns only the CMB-scale normalization and tilt. Whether the corresponding spectra remain viable at higher frequencies depends on their transfer history, spectral breaks and cutoffs. Turning these Fisher estimates into realistic detection forecasts will require full likelihood evaluations with realistic treatments of multifrequency foreground cleaning and systematic effects, which we leave for future work.

\begin{acknowledgments}
We thank Francesco Piacentini for access to CINECA computational resources through the INF26\_lspe project, and Umberto Ferraro Petrillo of the Department of Statistical Sciences, Sapienza University of Rome, for access to the Terastat2 HPC cluster. We also thank colleagues at the Dipartimento di Fisica, Sapienza Universit\`a di Roma, for useful discussions. The authors are supported by ``Theoretical Astroparticle Physics'' (TAsP), iniziativa specifica INFN. During the preparation of this work, the authors used the AI technologies Claude and Gemini 3.5 to polish the manuscript. All scientific ideas, analysis and interpretation of the results are the work of the authors, who take full responsibility for the final text.
\end{acknowledgments}

\appendix
\section{Joint likelihood profiling and confidence regions}\label{app:statistics}

Assuming the scalar parameters have been marginalized over, the remaining information on the tensor amplitude and tilt is described by the $2\times 2$ covariance matrix with off-diagonal correlation $\rho$. For a regular, local Gaussian model centered on the fiducial parameters, the likelihood displacement is
\begin{equation}
\Delta\chi^2=\frac{1}{1-\rho^2}
\left[\frac{\delta r^2}{\sigma_r^2}
-\frac{2\rho\,\delta r\,\delta n_t}{\sigma_r\sigma_{n_t}}
+\frac{\delta n_t^2}{\sigma_{n_t}^2}\right].
\label{eq:jointgaussian}
\end{equation}

To rigorously test the significance of a blue tilt, we evaluate the scale-invariant null hypothesis ($n_t=0$) while allowing the amplitude to vary freely. The profile-likelihood test statistic is
\begin{equation}
q_0=-2\ln\!\left[
\frac{\max_{r}\mathcal L(r,0)}
{\max_{r,n_t}\mathcal L(r,n_t)}
\right].
\label{eq:q0}
\end{equation}
Within the quadratic approximation, we enforce the null hypothesis by setting $\delta n_t=-n_{t,\rm fid}$ and minimize Eq.~\eqref{eq:jointgaussian} with respect to $\delta r$. The optimal amplitude shift that compensates for forcing $n_t=0$ is $\delta r_{\rm prof}=-\rho(\sigma_r/\sigma_{n_t})n_{t,\rm fid}$.
Substituting this back into the joint likelihood displacement, the correlation terms perfectly cancel the $(1-\rho^2)$ penalty prefactor. The resulting minimum displacement simplifies exactly to
\begin{equation}
q_0=\frac{n_{t,\rm fid}^2}{\sigma_{n_t}^2}=\Snt^2,
\label{eq:profile}
\end{equation}

showing that, in a purely Gaussian regime, the simple marginalized significance $\Snt$ is mathematically identical to a full likelihood-ratio test against scale invariance. Under standard asymptotic conditions \cite{Cowan2011}, the expected significance is $Z \simeq \Snt$, meaning our operational threshold of $\Snt=2$ corresponds to excluding $n_t=0$ at approximately $95.4\%$ confidence.

It is important to distinguish these independent marginal thresholds from a joint two-dimensional confidence region. A standard two-parameter Gaussian confidence ellipse, defined by $\Delta\chi^2 \leq \chi^2_{2;p}$, requires higher local significance levels to enclose the same probability volume due to the added degree of freedom. For instance, requiring a $95\%$ joint ellipse ($\Delta\chi^2 \leq 5.99$) to lie entirely within the physical, blue-tilted quadrant ($r>0, n_t>0$) would demand that both marginal ratios exceed $\sqrt{5.99} \simeq 2.45$. Our simultaneous criteria $\Sr>3$ and $\Snt>2$ therefore define an operational box for characterization rather than a formal joint probability contour.

Beyond defining this selection, the amplitude cut $\Sr>3$ serves a statistical purpose: it ensures that both the fiducial model and its optimal profiled shift $\delta r_{\rm prof}$ remain safely buffered from the $r=0$ boundary, where the tilt loses identifiability and the local Gaussian approximation breaks.

\section{Foreground model}\label{app:foregrounds}
\begin{figure}[h]
\centering
\includegraphics[width=\columnwidth]{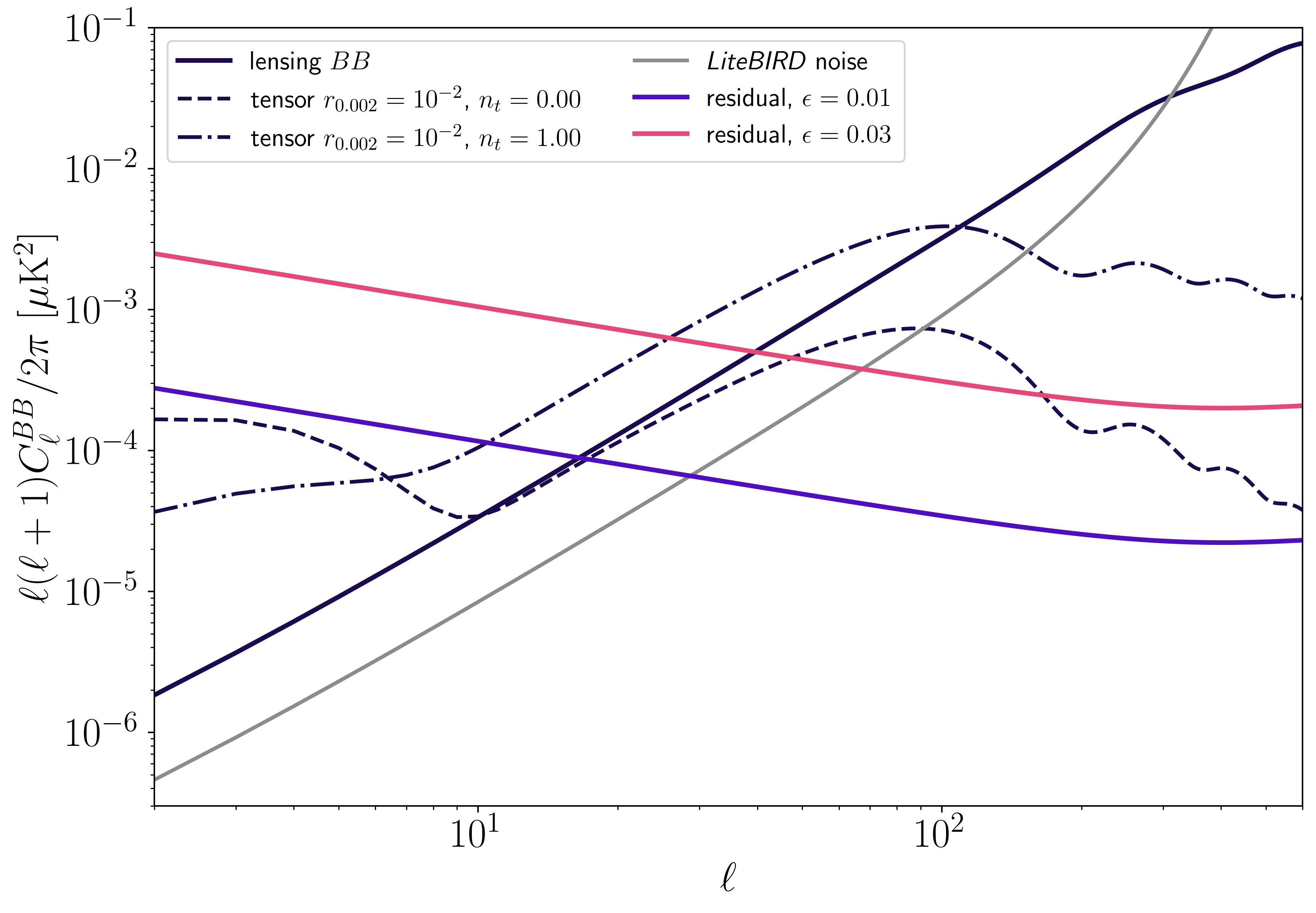}
\caption{Foreground residual power in the co-added map for $\epsilon=0.01$ and $0.03$, compared with the lensing $B$ modes, instrumental noise, and the tensor $B$ modes for $r_{0.002}=10^{-2}$ with $n_t=0$ and $n_t=1$.}\label{fig:fgres}
\end{figure}

The spectral energy distributions in thermodynamic CMB units, normalized at the reference frequency, are
\begin{equation}
\begin{split}
a_{\rm d}(\nu)&=\left(\frac{\nu}{\nu_{\rm d}}\right)^{\beta_{\rm d}+1}\frac{e^{h\nu_{\rm d}/k_BT_{\rm d}}-1}{e^{h\nu/k_BT_{\rm d}}-1}\,\frac{g(\nu)}{g(\nu_{\rm d})},\\
a_{\rm s}(\nu)&=\left(\frac{\nu}{\nu_{\rm s}}\right)^{\beta_{\rm s}}\frac{g(\nu)}{g(\nu_{\rm s})},
\end{split}
\end{equation}
with $g(\nu)=(e^{y}-1)^2/(y^2e^{y})$, $y=h\nu/k_BT_{\rm CMB}$ being the conversion from Rayleigh-Jeans to CMB temperature. The angular spectra at the reference frequencies are power laws, $\mathcal D_\ell^c=\ell(\ell+1)C_\ell^c/2\pi=\mathcal D_{80}^c(\ell/80)^{\gamma_c}$, with the parameters of Table~\ref{tab:fg}. For dust we use the large region LR71 of \textit{Planck} 2018 XI ($f_{\rm sky}\simeq0.7$, the sky fraction of our forecast) \cite{Planck2018XI}, while for synchrotron we use the \textit{Planck} 30\,GHz fit of Ref.~\cite{Martire2022} for their $94\%$ mask. We assume spatially constant spectral parameters and no dust-synchrotron correlation. Figure~\ref{fig:fgres} compares the residuals of Eq.~\eqref{eq:fgres} with the lensing and tensor $B$ modes.

As a check of the residual model, we repeated the calculation with the prescription of Refs.~\cite{Verde2006,Creminelli2015}, where each of the fifteen \textit{LiteBIRD} bands keeps a fraction $\epsilon^2$ of the foreground power, plus the noise of the foreground templates built from the lowest and highest bands. These residuals are treated as uncorrelated between channels, and all bands are combined with inverse-variance weights that include them. The characterized fraction becomes $34.4\%$ and $32.2\%$ for $\epsilon=0.01$ and $0.03$, against $30.7\%$ and $29.0\%$ with the model of Sec.~\ref{sec:fg} (Table~\ref{tab:robust}), and the onset stays at $n_t\simeq0.17$ and $0.18$. With the SO-like reconstruction combined with the CIB, the fractions are $39.7\%$ and $37.5\%$, against our $36.2\%$ and $34.2\%$. Their prescription is more optimistic because independent residuals in fifteen channels partly cancel when the channels are combined, whereas a residual left by the cleaning is the same sky in every channel, as Eq.~\eqref{eq:fgres} assumes, so we keep our model as the conservative choice.

\begin{table}[h]
\caption{Foreground models: $\mathcal D_{80}$ is the $BB$ amplitude at $\ell=80$ at the reference frequency and $\gamma$ the slope of $\mathcal D_\ell$.}\label{tab:fg}
\begin{tabular}{lcccccc}
\hline\hline
 & $\nu_{\rm ref}$ [GHz] & $\mathcal D_{80}$ [$\mu$K$^2$] & $\gamma$ & $\beta$ & $T_{\rm d}$ [K] & Ref.\\
\hline
Dust & 353 & 167 & $-0.54$ & 1.53 & 19.6 & \cite{Planck2018XI}\\
Synchrotron & 28.4 & 1.67 & $-0.22$ & $-3.05$ & & \cite{Martire2022}\\
\hline\hline
\end{tabular}
\end{table}

\section{Delensing with specific tracers}\label{app:delensing}
\begin{figure*}[!th]
\centering
\includegraphics[width=0.94\textwidth]{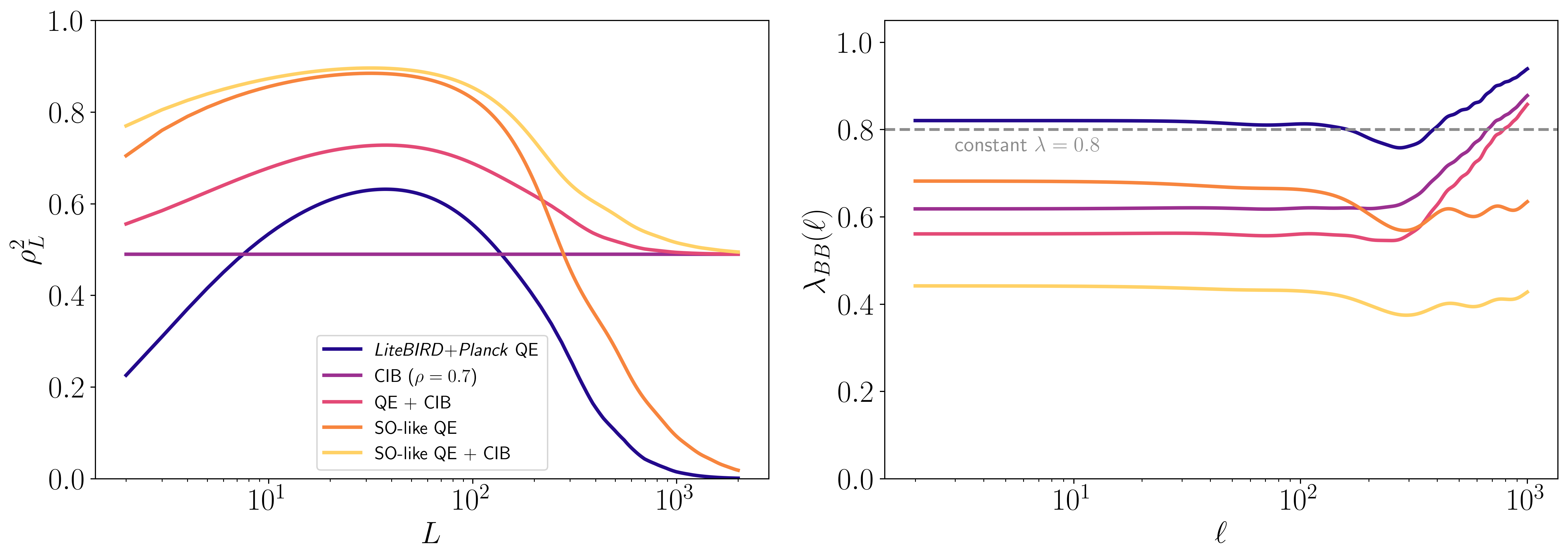}
\caption{Left: squared correlation coefficient between each tracer and the lensing potential. Right: fraction of lensing $B$-mode power left after delensing with that tracer, compared with the constant benchmark $\lambda=0.8$ (dashed).}\label{fig:tracers}
\end{figure*}

Delensing with a specific tracer of the lensing potential $\phi$ \cite{KnoxSong2002,Kesden2002} gives an $\ell$-dependent residual, which replaces the constant $\lambda$ of Eq.~\eqref{eq:delensing} with $\lambda_\ell$. If the tracer has correlation coefficient $\rho_L$ with $\phi$ and the $E$ modes are Wiener filtered with $W^E_\ell=C_\ell^{EE}/(C_\ell^{EE}+N_\ell^P)$, the residual lensing power is \cite{Smith2012,SherwinSchmittfull2015}
\begin{equation}
C_\ell^{BB,\rm res}=\mathcal B_\ell\big[C^{EE},C^{\phi\phi}\big]-\mathcal B_\ell\big[W^EC^{EE},\rho^2C^{\phi\phi}\big],
\end{equation}
where $\mathcal B_\ell[\cdot,\cdot]$ is the lensing $B$-mode spectrum generated by a given $E$-mode and lensing-potential spectrum, which we compute with \texttt{CAMB} \cite{LewisChallinor2006}. The ratio $\lambda_\ell=C_\ell^{BB,\rm res}/(C_{\ell,\rm len}^{BB}-C_{\ell,\rm unl}^{BB})$ then enters Eq.~\eqref{eq:delensing} as a multipole-dependent residual fraction, which reduces to $\lambda=1-\rho^2$ for a constant $\rho$ and noiseless $E$ modes. For a quadratic estimator, the reconstructed field is $\phi$ plus a reconstruction noise $N_L^{\phi\phi}$ uncorrelated with it, so that the correlation is $\rho_L^2=C_L^{\phi\phi}/(C_L^{\phi\phi}+N_L^{\phi\phi})$ \cite{HuOkamoto2002,OkamotoHu2003}. For the internal reconstruction we compute $N_L^{\phi\phi}$ in the flat-sky approximation from the $TT$, $EE$ and $EB$ estimators, combined in minimum variance, with the lensed spectra and the noise of Table~\ref{tab:noise}. The cosmic infrared background (CIB) enters only through its correlation with $\phi$, which we take constant for simplicity, $\rho=0.7$ \cite{Planck2013_XVIII,SherwinSchmittfull2015,Yu2017}. When several tracers are combined, the optimal combination has $\rho^2=\bm c^{\rm T}\mathsf C^{-1}\bm c/C^{\phi\phi}$, with $\bm c$ the tracer-$\phi$ cross spectra and $\mathsf C$ the tracer covariance \cite{SherwinSchmittfull2015}. For a quadratic estimator combined with the CIB we normalize both tracers to unit variance, so that $\bm c^{\rm T}=\sqrt{C^{\phi\phi}}\,(\rho_{\rm QE},\rho_{\rm CIB})$. The reconstruction noise is uncorrelated with the CIB, so the two tracers are correlated only through $\phi$ and the off-diagonal element of $\mathsf C$ is $\rho_{\rm QE}\rho_{\rm CIB}$. For the Simons Observatory, we take the public noise curves of its large-aperture telescope at baseline sensitivity, for $f_{\rm sky}=0.4$ and without foreground deprojection \cite{SimonsObs2019},\footnote{\url{https://github.com/simonsobs/so_noise_models}} using the minimum-variance lensing reconstruction noise for $\phi$ and the component-separated $E$-mode noise for the Wiener filter of the template. For the other tracers, the $E$ modes come from \textit{LiteBIRD} and \textit{Planck}. We consider the internal reconstruction, the CIB, the Simons Observatory reconstruction, and the combination of each estimator with the CIB. Figure~\ref{fig:tracers} shows the resulting $\rho_L^2$ and $\lambda_\ell$. The residual fraction stays nearly constant up to $\ell\sim100$, since lensing $B$ modes on those scales are produced by lenses at $L\sim100$ to $1000$, and at higher $\ell$ it eventually rises due to $E$ modes becoming noisy. At $\ell=80$, internal reconstruction leaves $\lambda_{80}=0.81$, the value that motivates our benchmark, while the CIB gives $0.62$, internal reconstruction combined with the CIB $0.56$, the SO-like estimator $0.66$, and its combination with the CIB $0.43$.

\bibliographystyle{apsrev4-2}
\bibliography{refs_draft}

\end{document}